\documentclass[conference]{IEEEtran}
\usepackage{stfloats}
\usepackage{xcolor}
\usepackage{colortbl}
\usepackage[switch]{lineno}

\colorlet{shadecolor}{gray!40}
\definecolor{apricot}{rgb}{0.98, 0.81, 0.69}
\UseRawInputEncoding
\makeatletter
\newcommand\notsotiny{\@setfontsize\notsotiny{6}{7}}

\makeatother

\usepackage{url}

\usepackage{listings}
\usepackage{xcolor}
\usepackage{listings}

\definecolor{mGreen}{rgb}{0,0.6,0}
\definecolor{mGray}{rgb}{0.5,0.5,0.5}
\definecolor{mPurple}{rgb}{0.58,0,0.82}
\definecolor{backgroundColour}{rgb}{0.95,0.95,0.92}

\lstdefinestyle{CStyle}{
    backgroundcolor=\color{backgroundColour},   
    commentstyle=\color{mGreen},
    keywordstyle=\color{magenta},
    numberstyle=\tiny\color{mGray},
    stringstyle=\color{mPurple},
    basicstyle=\notsotiny,
    breakatwhitespace=false,         
    breaklines=true,                 
    captionpos=b,                    
    keepspaces=true,                 
    numbers=left,                    
    numbersep=5pt,                  
    showspaces=false,                
    showstringspaces=false,
    showtabs=false,
    tabsize=2,
    language=C,
    moredelim=**[is][\color{blue}]{~}{~},
}

\usepackage{xspace}
\usepackage{placeins}
\usepackage{cite}

\usepackage{algorithm}
\usepackage[noend]{algpseudocode}
\algnewcommand{\LineComment}[1]{\State \emph{\textcolor{blue}{\(\triangleright\) #1}}}
\algrenewcommand\algorithmicindent{1em}%

\usepackage{graphicx}
\usepackage{caption}
\usepackage{subcaption}
\usepackage{tabularx}
\usepackage{enumitem}   

\usepackage{color, soul} 
\soulregister\emph7
\soulregister\cite7
\soulregister\ref7
\soulregister\pageref7
\definecolor{Gray}{gray}{0.85}
\definecolor{LightCyan}{rgb}{0.88,1,1}

\begin{document}

\newfloat{lstfloat}{htbp}{lop}
\floatname{lstfloat}{Listing}
\def\lstfloatautorefname{Listing} % needed for hyperref/auroref

\setlength{\abovedisplayskip}{-1pt}
\setlength{\belowdisplayskip}{-1pt}

\setlength\floatsep{0.3\baselineskip plus 1pt minus 20pt}
\setlength\textfloatsep{0.3\baselineskip plus 1pt minus 20pt}
\setlength\intextsep{0.3\baselineskip plus 1pt minus 20pt}
\newcolumntype{R}{>{\centering\arraybackslash}m{3.5cm}}
\newcolumntype{L}{>{\centering\arraybackslash}m{1.5cm}}
\newcolumntype{M}{>{\centering\arraybackslash}m{3.5cm}}

\title{Harnessing Deep Learning and HPC Kernels via High-Level Loop and Tensor Abstractions on CPU Architectures}

\author{\IEEEauthorblockN{Evangelos Georganas\IEEEauthorrefmark{1},
Dhiraj Kalamkar\IEEEauthorrefmark{1}, Kirill Voronin\IEEEauthorrefmark{1}\textsuperscript{\textsection},
Abhisek Kundu\IEEEauthorrefmark{1},
Antonio Noack\IEEEauthorrefmark{2},\\
Hans Pabst\IEEEauthorrefmark{1},
Alexander Breuer\IEEEauthorrefmark{2},
and
Alexander Heinecke\IEEEauthorrefmark{1}}
\IEEEauthorblockA{\IEEEauthorrefmark{1}Intel Corporation\\
\IEEEauthorrefmark{2}Friedrich Schiller Universität Jena}}

\maketitle
\begingroup\renewcommand\thefootnote{\textsection}
\footnotetext{Now at NVIDIA Corporation}
\endgroup

\begin{abstract}
During the past decade, Deep Learning (DL) algorithms, programming systems and hardware have converged with the High Performance Computing (HPC) counterparts. Nevertheless, the programming methodology of DL and HPC systems is stagnant, relying on highly-optimized, yet platform-specific and inflexible vendor-optimized libraries. Such libraries provide close-to-peak performance on specific platforms, kernels and shapes thereof that vendors have dedicated optimizations efforts, while they underperform in the remaining use-cases, yielding non-portable codes with performance glass-jaws. This work introduces a framework to develop efficient, portable DL and HPC kernels for modern CPU architectures. We decompose the kernel development in two steps: 1) Expressing the computational core using Tensor Processing Primitives (TPPs): a compact, versatile set of 2D-tensor operators, 2) Expressing the logical loops around TPPs in a high-level, declarative fashion whereas the exact instantiation (ordering, tiling, parallelization) is determined via simple knobs. We demonstrate the efficacy of our approach using standalone kernels and end-to-end workloads that outperform state-of-the-art implementations on diverse CPU platforms.
\end{abstract}

\section{Introduction}
\label{sec:introduction}
Deep Learning (DL) emerged as a promising machine learning paradigm more than a decade ago, and since then, deep neural networks have made significant advancements in various fields, including computer vision, natural language processing, and recommender systems, while gradually expanding their applications in traditional scientific domains~\cite{origalexnet,szegedy2015going,simonyan2014very,yu2013feature,wu2016google,cheng2016wide,wolf2020transformers,gawehn2016deep,goh2017deep,raghu2020survey}. Despite the apparent differences between conventional High Performance Computing (HPC) and DL workloads, the computational kernels used in both fields largely overlap. These kernels involve tensor contractions (dense and sparse), elementwise tensor operations, tensor norm computations, and generalized tensor re-orderings~\cite{di2022high,georganas2021tensor}.

The deployment of programming systems, algorithms, and hardware in the domain of DL has led to convergence with the HPC counterparts. However, the prevalent programming paradigm has become stagnant compared to the rapidly evolving DL/HPC workloads. The programming paradigm relies heavily on vendor-optimized libraries for vital building blocks of applications. These libraries offer close-to-peak performance on specific platforms, kernels, and shapes thereof that vendors have dedicated optimization efforts. However, they under-perform in other use-cases, resulting in non-portable, inflexible codes with performance limitations. Meanwhile, mainstream CPU platforms have continued to evolve, offering hardware acceleration capabilities for key operations, core heterogeneity, large core-counts, and complex memory hierarchies. Such nuances of contemporary CPU platforms make it impractical to use ``one-fits-all" code generation strategies, exacerbating the generalization problem inherited by extended library dependencies. Consequently, this complexity is transferred to library development, and libraries for key computations become highly specialized, making them difficult to develop, maintain, and extend to encompass the latest CPU advances~\cite{barham2019machine}. In the quest of eliminating the dependency from vendor-libraries, the research fields of Tensor Compilers (e.g.~\cite{plaidml,chen2018tvm,vasilache2018tensor,zheng2020ansor}) and Domain Specific Languages (DSLs) (e.g.~\cite{zhang2017snowflake, ragan2013halide}) are experiencing a renaissance. Nevertheless, such efforts typically provide point solutions (often domain specific), and have not proven their applicability on real-world use-case scenarios~\cite{barham2019machine}.

 We observe that the root cause of the aforementioned problems can be traced back to the extreme abstraction levels provided by libraries and tensor compilers. The libraries are characterized by coarse-grain, monolithic kernels that lack flexibility, while compilers allow for unrestricted expression of low-level operators, which can hinder efficient code generation in the back-ends. Adding to the difficulty of achieving optimal code generation, tensor compilers are responsible for efficiently parallelizing, tiling, and re-ordering loops, as well as transforming the layout of data structures. All these tasks remain unsolved in a generic setup to date. These observations suggest that a middle path that embraces high-level abstractions for the loop generation and the tensor operations, and emphasizes clear separation of concerns can effectively tackle the majority of the challenges~\cite{barham2019machine}.

In this study we built upon and extend the prior work of Georganas et.al.~\cite{georganas2021tensor}, namely Tensor Processing Primitives (TPP). TPPs comprise a collection of basic operators on 2D tensors that can be used to construct more complex operations on high-dimensional tensors. The TPP collection is compact, expressive, and precision-aware, and as a result high-level DL and HPC kernels can be written in terms of TPP operations. The specification for TPPs is agnostic to platform, DL framework, and compiler back-end, which makes the TPP-code portable. TPPs operate at the sub-tensor granularity, making them directly accessible to workload and library developers. While the TPP specification is platform-agnostic, its implementation is platform-specific, and optimized for the target architectures. This separation of concerns allows the user-entity of TPPs to focus on the algorithm and its execution schedules, while the TPP back-end generates efficient, platform-specific code for the TPP operations.

While the TPP abstraction addresses the code generation problem of the core computation kernels, it is still the user-entity's responsibility to write the required nested loops around the TPP primitives (henceforth called ``outer loops'') that essentially traverse the computation/iteration space of the kernel. These ``outer loops'' effectively control the parallelization scheme of the computation, and the corresponding loop orders in conjunction with potential loop blockings/tilings affect the temporal and spatial locality of the computation~\cite{sc18,georganas2020harnessing}. Consequently, in order to get high-performance kernels, the user has to write (potentially) complicated, non-portable (across platforms) code with respect to these ``outer loops'' that takes into account the increasingly complex memory hierarchy of the diverse compute platform, and the increased degree of available parallelism. Also, the exact instantiation of these loops that yields optimal kernel performance directly depends on the problem size and the platform at hand (i.e.\ there is no ``one-fits-all'' solution, an issue that also hinders portable library performance).

\begin{figure}[t!]
\centering
\includegraphics[width=1.04\columnwidth]{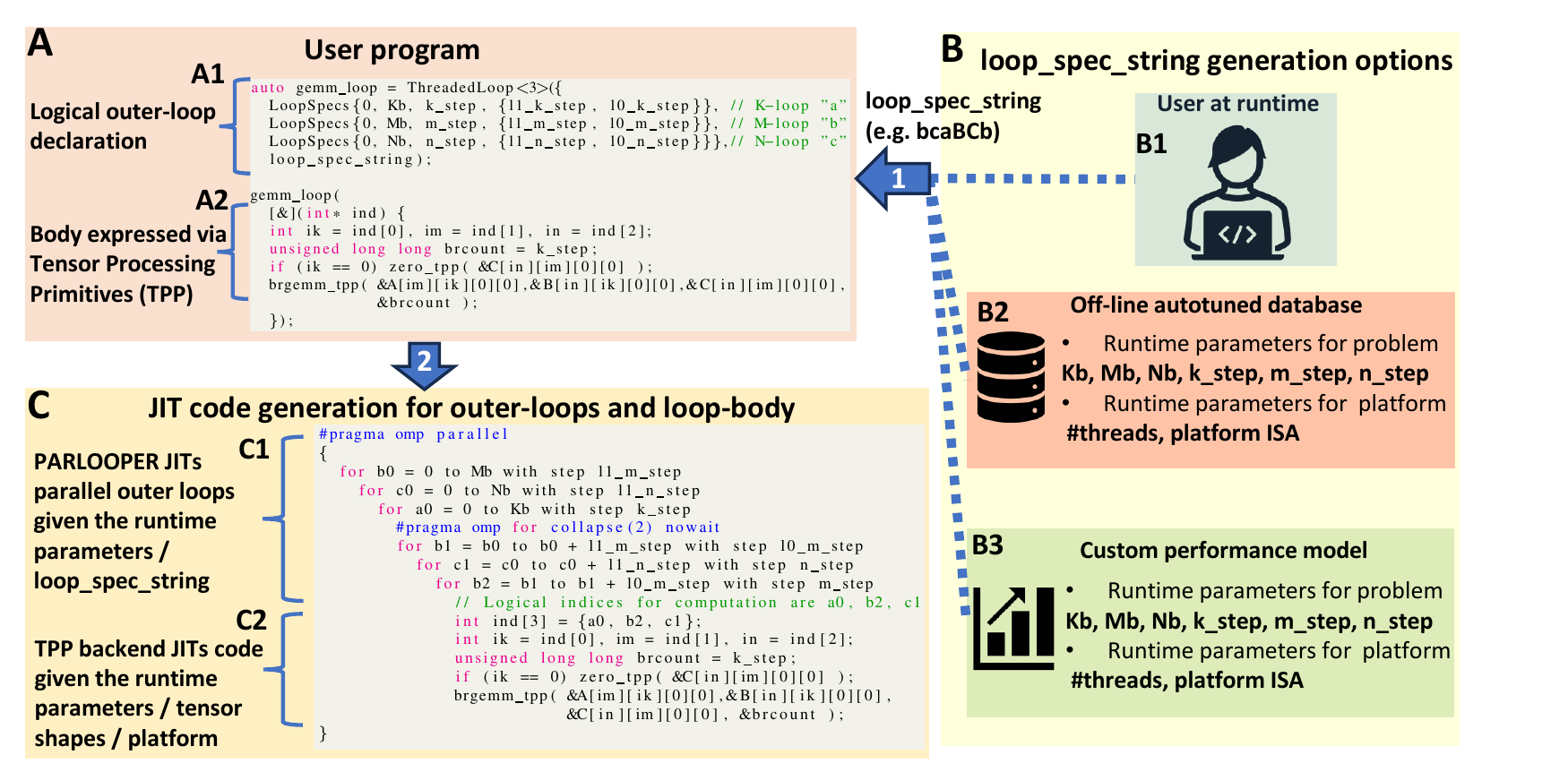}
\caption{PARLOOPER \& TPP framework workflow}
\label{fig:workflow}
\end{figure}

We address the problem of generating arbitrarily complex parallel loops around TPPs by introducing PARLOOPER (PARallel LOOP gEneratoR), a high-level loop abstraction framework (see Figure~\ref{fig:workflow}). PARLOOPER aims to simplify the ``outer loop'' writing (Figure~\ref{fig:workflow}-A1) by enabling the user to declare the logical ``outer loops'' along with their specifications (i.e. bounds/steps/parallelization properties), instead of explicitly writing the tedious loop nests that pertain to multiple loop orders, tilings and parallelization schemes. At runtime, the user may provide a single parameter (henceforth called \emph{loop\_spec\_string}, Figure~\ref{fig:workflow}-Arrow 1) to dictate the desired instantiation of the loop nest (i.e. loop order, loop blockings/tilings, parallelization method). In our Proof-Of-Concept (POC) implementation, PARLOOPER is a lightweight C++ library that auto-generates Just-In-Time (JIT) the requested instantiation of the loop nest by considering the loop specifications and the single \emph{loop\_spec\_string} runtime parameter (Figure~\ref{fig:workflow}-Arrow 2 and Box C1). The resulting user code (Figure~\ref{fig:workflow}-Box A) is extremely simple, declarative, and high-level yet powerful: With zero lines of code-changes the loop nest can be instantiated to arbitrarily complex implementation that maximizes the performance for a specific platform and problem at hand. PARLOOPER uses internally caching schemes to avoid JIT overheads whenever possible. By leveraging PARLOOPER along with the TPP programming abstraction for the core computation, the user code is simple/declarative, and naturally lends itself to auto-tuning to explore complex outer loop instantiations with zero lines of user-code writing. %This allows for efficient implementation of loop nests/tensor computations with skewed sizes whereas libraries usually ill-behave.

We decompose the kernel development in two steps with clear separation of concerns. First, we express the computational core/body of a kernel using solely Tensor Processing Primitives and the \emph{logical} loop indices of the surrounding/``outer'' loops (Figure~\ref{fig:workflow}-Box A2). Second, we declare the surrounding \emph{logical} loops along with their specification (i.e. bounds/steps/parallelization properties) with PARLOOPER (Figure~\ref{fig:workflow}-Box A1). The result is a streamlined, compact code that is identical for all target platforms/problem sizes/shapes; the exact instantiation of the loop-nest (Figure~\ref{fig:workflow}-Box C1) is governed by a simple runtime knob, whereas the efficient code generation pertaining to TPP (incl. loop unrolling, vectorization, register blocking, instruction selection) is undertaken by the TPP back-end which is JITing code based on the target platform (Figure~\ref{fig:workflow}-Box C2)~\cite{georganas2021tensor}.

The \emph{loop\_spec\_string knob} in PARLOOPER can be determined either ``manually" by the user (i.e. Figure~\ref{fig:workflow}-Box B1 the user selects values for the knob that yield good performance based on custom performance modeling) or by auto-tuning techniques (Figure~\ref{fig:workflow}-Box B2), where a bunch of values for the \emph{loop\_spec\_string knob} are benchmarked offline and the best one is selected during runtime (Figure~\ref{fig:workflow}-Arrow 1). To assist the selection of the \emph{loop\_spec\_string} runtime knob, instead of relying exclusively on manual settings/exhaustive auto-tuning, we provide an exemplary lightweight, high-level performance modeling tool (Figure~\ref{fig:workflow}-Box B3).

We demonstrate the efficiency of our approach on multiple platforms using standalone kernels and end-to-end workloads, and compare the performance to State-Of-The-Art implementation. The main contributions of this work are:

 \begin{itemize}[labelindent=0em,labelsep=0.2cm,leftmargin=*,noitemsep,topsep=0pt]
 \item A lightweight, flexible and high-level parallel loop generation framework (PARLOOPER) that decouples the logical loop declaration from the nested loop instantiation. The user-code is declaring the logical loops at high-level, whereas the exact instantiation is controlled by one knob.
 \item Extending the LIBXSMM TPP backend~\cite{georganas2021tensor} to support the ARM SVE ISA and hardware accelerated tensor-contractions on the Graviton 3 platform. Also, we introduce sparse$\times$dense matrix multiplication TPPs with block sparsity, low-precision support and hardware acceleration for both x86 and ARM/AArch64 platforms.
 \item Exemplary DL and HPC multi-threaded kernels using the PARLOOPER framework and the TPP abstraction for the core computations (GEMM, convolutions, MLP, block-sparse$\times$dense GEMM) including multiple precisions.% (FP32, BF16~\cite{bfloat16_tf}, FP8~\cite{micikevicius2022fp8}, INT8).
 \item Implementing end-to-end DL workloads via the PARLOOPER/TPP paradigm for Image Recognition (ResNet50~\cite{he2016deep}) and Large Language Model pipelines.
 \item Benchmarking the standalone kernels and end-to-end workloads on multiple modern CPUs with various datatype precisions while matching/exceeding state-of-the-art (SOTA) performance from vendor libraries.
 \end{itemize}

\section{The PARLOOPER Framework}
\label{sec:parlooper_spec}
\subsection{GEMM Example Notations}
We illustrate our PARLOOPER/TPP framework with a running General Matrix Multiplication (GEMM) example (see Figure~\ref{fig:workflow} and Listing~\ref{lst:gemm}). The \emph{zero\_tpp} sets the input 2D tensor to zero values. The \emph{brgemm\_tpp} corresponds to the \emph{Batch-Reduce GEMM} (BRGEMM) TPP which is the main tensor contraction tool in the TPP collection~\cite{georganas2020harnessing,georganas2021tensor}. BRGEMM materializes $C = \beta \cdot C + \sum_{i=0}^{brcount-1} A_i \times B_i$. This kernel multiplies the specified blocks $A_i^{bm\times bk}$ and $B_i^{bk\times bn}$ and reduces the partial results to a block $C^{bm\times bn}$. We use the \emph{stride-based} variant of BRGEMM, where the addresses of $A_i$ and $B_i$ are: $address\_A_i = address\_A_{i-1} + stride\_A$ and $address\_B_i = address\_B_{i-1} + stride\_B$~\cite{georganas2021tensor}. The Matrix Multiplication input tensors are logically 2D matrices $A^{M\times K}$ and $B^{K\times N}$ that need to be multiplied and added onto $C^{M\times N}$. We follow the approach of previous work~\cite{georganas2020harnessing} and we block the dimensions $M$, $K$, and $N$ by factors $bm$, $bk$, and $bn$ respectively (lines 1-3 of Listing~\ref{lst:gemm}). We employ the BRGEMM TPP to perform the tensor contraction with $A$ and $B$ across their dimensions $Kb$ and $bk$. The sub-blocks $A_i$ and $B_i$ that have to be multiplied and reduced are apart by fixed strides $stride\_A=bk\cdot bm$ and $stride\_B=bn\cdot bk$.

\subsection{Logical Loop Declaration}
\label{subsec:loop_decl}
\begin{lstfloat}[t]
\begin{lstlisting}[style=CStyle]
DType A[Mb][Kb][bk][bm]; //Block M dim with bm and K dim with bk
DType B[Nb][Kb][bn][bk]; //Block N dim with bn and K dim with bk
DType C[Nb][Mb][bn][bm]; //Block N dim with bn and M dim with bm

auto gemm_loop = ThreadedLoop<3>({
  LoopSpecs{0, Kb, k_step, {l1_k_step, l0_k_step}}, // K-loop "a"
  LoopSpecs{0, Mb, m_step, {l1_m_step, l0_m_step}}, // M-loop "b"
  LoopSpecs{0, Nb, n_step, {l1_n_step, l0_n_step}}},// N-loop "c"
  loop_spec_string);
  
gemm_loop(
  [&](int* ind) {
  int ik = ind[0], im = ind[1], in = ind[2];
  unsigned long long brcount = k_step;
  if (ik == 0) zero_tpp( &C[in][im][0][0] );
  brgemm_tpp( &A[im][ik][0][0],&B[in][ik][0][0],&C[in][im][0][0],
              &brcount );
  });
\end{lstlisting}
%\vspace{-1.5em}
\caption{GEMM written with PARLOOPER and TPPs}
\label{lst:gemm}
\end{lstfloat}
The Matrix Multiplication algorithm is comprised of three logical nested loops which are declared in lines 5-9 of Listing~\ref{lst:gemm}/Figure~\ref{fig:workflow}-Box A1. The computation involves three logical loops, thus we declare: ThreadedLoop$<3>$ (line 5).

The first loop (line 6) with mnemonic \emph{a}, corresponds to a loop with start $0$, upper bound $Kb$ and step $k\_step$, and it reflects the ``$K$" loop of the GEMM ($K$ is the inner-product dimension). This loop has an optional list of step/blocking parameters $\{l1\_k\_step,\ l0\_k\_step\}$. The second loop (line 7) with the mnemonic \emph{b}, corresponds to a loop with start $0$, upper bound $Mb$ and step $m\_step$, and reflects the ``$M$" loop of the GEMM. Similarly, this loop has an optional list of step/blocking parameters $\{l1\_m\_step,\ l0\_m\_step\}$. Finally, the third loop (line 8) with mnemonic \emph{c}, corresponds to a loop with start $0$, upper bound $Nb$ and step $n\_step$, and reflects the ``$N$" loop of the GEMM. This loop has an optional list of step/blocking parameters $\{l1\_n\_step,\ l0\_n\_step\}$.

\begin{lstfloat}[t!]
\begin{lstlisting}[style=CStyle]
void parlooper_nested_loops(std::function<void(int*)> body_func,
                            std::function<void()>     init_func,
                            std::function<void()>     term_func) {
  ~#pragma omp parallel~
  {
    if (init_func) init_func();
    for b0 = 0 to Mb with step l1_m_step
      for c0 = 0 to Nb with step l1_n_step
        for a0 = 0 to Kb with step k_step
          ~#pragma omp for collapse(2) nowait~
          for b1 = b0 to b0 + l1_m_step with step l0_m_step
            for c1 = c0 to c0 + l1_n_step with step n_step
              for b2 = b1 to b1 + l0_m_step with step m_step
                // Logical indices for computation are a0, b2, c1
                int ind[3] = {a0, b2, c1};
                body_func(ind);
    if (term_func) term_func();
  }
}
\end{lstlisting}
%\vspace{-1.5em}
\caption{Generated loop for $loop\_spec\_string$=bcaBCb}
\label{lst:collapse_loop}
\end{lstfloat}

The specific instantiation of these loops, i.e. the loop order with which they appear, the number of times each one is blocked and also the way they are parallelized are controlled by the run-time knob \emph{loop\_spec\_string} (line 9). Given such a string during run-time, the constructor \emph{ThreadedLoop} invokes our custom loop generator and emits a C++ function for the target loop instantiation (Figure~\ref{fig:workflow}-Box C1). Listing~\ref{lst:collapse_loop} illustrates an exemplary target-loop instantiation function. Subsequently, a C++ compiler is invoked (currently we support icc, clang, and gcc), and the target-loop instantiation function is compiled Just-In-Time (JIT), which can be further used by the code (line 11 in Listing~\ref{lst:gemm}). To avoid JIT overheads whenever possible, we cache the JITed target loops: if we request a loop nest with the same \emph{loop\_spec\_string}, we merely return the function pointer of the already compiled and cached loop-nest.% Section~\ref{subsec:computation} provides more details with respect to the target-loop instantiation function and its API.

In regard to the legality/validity of the \emph{loop\_spec\_string} (i.e. what loop permutations are allowed), this is responsibility of the user entity and depends on the computation at hand. The two rules for constructing a valid \emph{loop\_spec\_string} are:
\begin{itemize}[labelindent=0em,labelsep=0.0cm,leftmargin=*,noitemsep,topsep=0pt]
\item\textbf{\ RULE 1: Loop ordering and blockings}
\end{itemize}

Each character (from \emph{a} to \emph{z}, depending on the number of the logical loops) can appear in any order and any number of times. In our case, since we have 3 logical loops the characters range from $a$ to $c$. The order with which the loop characters appear in the string determine the nesting loop order, and the times each character appears determines how many times the corresponding logical loop is blocked. For example, a \emph{loop\_spec\_string} \emph{bcabcb} corresponds to a loop where logical loop $b$ is blocked twice (the character $b$ appears 3 times), logical loop $c$ is blocked once (the character $c$ appears 2 times) and the logical loop $a$ is not blocked (it appears only once). The blocking/tiling sizes for each logical loop level are extracted from the corresponding list of step/blocking parameters in order they appear in the list. Our Proof-Of-Concept (POC) implementation allows only perfectly nested blocking/tiling sizes, i.e.\ in the example above it should hold:
$l1\_m\_step\ \%\ l0\_m\_step = 0$, $l0\_m\_step\ \%\ m\_step = 0$ and $l1\_n\_step\ \%\ n\_step = 0$. All these blocking/tiling sizes lists may be provided at runtime (e.g., programmatically determine the blocking sizes given a problem) and do not have to be statically determined.

\begin{itemize}[labelindent=0em,labelsep=0.0cm,leftmargin=*,noitemsep,topsep=0pt]
\item\textbf{\ RULE 2: Parallelization of loops}
\end{itemize}

If a loop character in the \emph{loop\_spec\_string} appears in its upper-case form, it dictates the intention to parallelize this loop at the specific nesting level it appears. In the previous example, the \emph{loop\_spec\_string} \emph{bcaBcb}, corresponds to a loop nest where the 2nd occurrence of loop $b$ is parallelized. Our implementation supports 2 modes of parallelization:

\emph{\textbf{ PAR-MODE 1: Relying on OpenMP~\cite{chandra2001parallel} runtime for parallelizing the loops.}} In this method, the loop parallelization strategy corresponds to the directive\texttt{\footnotesize \#pragma omp for nowait}. If the user intends to parallelize multiple loops, the corresponding capitalized characters should appear consecutively in the \emph{loop\_spec\_string}, and it would result in parallelization using collapse semantics. E.g., for the \emph{loop\_spec\_string} $bcaBCb$, PARLOOPER generates the loop nest in Listing~\ref{lst:collapse_loop}. We allow optionally adding directives at the end of the \emph{loop\_spec\_string} by using the special character $@$ as separator. E.g., the loop string ``$bcaBCb$ $@$ $schedule(dynamic,1)$" yields the directive: \texttt{\footnotesize \#pragma omp for collapse(2) schedule(dynamic,1) nowait}. This method allows PARLOOPER to leverage features from the OpenMP runtime like dynamic thread scheduling. The constructed loop nest is embraced by a \texttt{\footnotesize\#pragma omp parallel} region (line 4 in Listing~\ref{lst:collapse_loop}). A barrier at the end of a specific loop-level may be requested using the special character ``$\vert$".

\emph{\textbf{\ PAR-MODE 2: Using explicit multi-dimensional thread decompositions}}. The user can specify explicit 1D, 2D or 3D loop parallelization schemes. For the 1D decomposition, the threads form a logical $R\times 1$ 1D grid and are assigned the corresponding parallelized loop iterations in a block fashion. The user merely has to append after the desired upper-case loop character the substring \emph{\{R:\#threads\}}, where $\#threads$ is a number dictating in how many ways to parallelize that loop. For a 2D decomposition, the threads are forming a logical $R\times C$ 2D grid that effectively parallelizes the requested two loops. E.g.,\ consider the \emph{loop\_spec\_string} \emph{bC\{R:16\}aB\{C:4\}cb}. The loop $c0$ is parallelized 16-ways and the loop $b1$ is parallelized 4-ways using a logical thread grid of 16$\times$4 ($R=16$, $C=4$) -- see Listing~\ref{lst:2d_decomp_loop}. Each loop that is parallelized is done so in a block fashion using the requested number of ``ways". The 3D decomposition works in an analogous manner.

The current POC supports OpenMP for concurrency purposes, however our back-end loop generator can be  extended to support other runtimes (e.g. TBB~\cite{pheatt2008intel} or pthreads~\cite{nichols1996pthreads}).

\begin{lstfloat}[t!]
\begin{lstlisting}[style=CStyle]
# pragma omp parallel
{
  int tid = omp_get_thread_num();
  int row_teams = 16, col_teams = 4;
  int row_id= id/col_teams, col_id = tid%col_teams;
  if (init_func) init_func();
  for b0 = 0 to Mb with step l1_m_step
  ~# Parallelize 16-ways loop c0, assign tasks based on row_id~
  for c0 = 0 to Nb with step l1_n_step
    for a0 = 0 to Kb with step k_step 
      ~# Parallelize 4-ways loop b1, assign tasks based on col_id~ 
      for b1 = b0 to b0 + l1_m_step with step l0_m_step
        for c1 = c0 to c0 + l1_n_step with step n_step
          for b2 = b1 to b1 + l0_m_step with step m_step
            // Logical indices for computation are a0, b2, c1
            int ind[3] = {a0, b2, c1};
            body_func(ind);
  if (term_func) term_func();
}
\end{lstlisting}
%\vspace{-1.5em}
\caption{Loop for $loop\_spec\_string$=bC\{R:16\}aB\{C:4\}cb}
\label{lst:2d_decomp_loop}
\end{lstfloat}

\subsection{Expressing the Computation}
\label{subsec:computation}
After declaring the nested loops, we obtain a ThreadedLoop object (\emph{gemm\_loop} in Line 5 of Listing~\ref{lst:gemm}) which can be passed at runtime (up to) three parameters:

\begin{itemize}[labelindent=0em,labelsep=0.0cm,leftmargin=*,noitemsep,topsep=0pt]
\item \ \emph{(Optional)} A pointer to a function: \textbf{ \emph{void init\_func()}}
\end{itemize}
This function is called before the generated loop-nest and is used for ``initialization" purposes (e.g., code that initializes data structures etc.).
\begin{itemize}[labelindent=0em,labelsep=0.0cm,leftmargin=*,noitemsep,topsep=0pt]
\item \ \emph{(Optional)} A pointer to a function: \textbf{ \emph{void term\_func()}}
\end{itemize}
This function is called after the generated loop-nest and is used for ``termination " purposes (e.g., code that cleans up data structures etc).
\begin{itemize}[labelindent=0em,labelsep=0.0cm,leftmargin=*,noitemsep,topsep=0pt]
\item \ A pointer to a function: \textbf{ \emph{void body\_func(int *ind)}}
\end{itemize}
The function \emph{body\_func} is called at the inner-most level of the generated loop-nest (e.g.\ line 16 in Listing~\ref{lst:collapse_loop}), and it performs the desired computation. The function \emph{body\_func} gets as input an array of integer values which contains the values of the logical indices used in the nested loop in alphabetical order: \emph{ind[0]} corresponds to the value of the logical index $a$ in the current nested-loop iteration, \emph{ind[1]} corresponds to the logical index $b$ in the current iteration etc. This index array is allocated and initialized by PARLOOPER (e.g.\ line 15 in Listing~\ref{lst:collapse_loop}). By leveraging the values of the logical indices, the user can express the desired computation as a function of those. We use as \emph{body\_func} an in-place C++ lambda expression, but this is not a requirement in our framework.
Considering the GEMM example, and using lambda expression for \emph{body\_func} we express the GEMM computation using the \emph{zero\_tpp}, the \emph{brgemm\_tpp}, and the logical indices with lines 12-17 of Listing~\ref{lst:gemm} and Figure~\ref{fig:workflow}-Box A2. In line 13, we extract the logical indices $ik$, $im$ and $in$ from the input array $ind$. We set batch-reduce count in Line 14 to $k\_step$ since the tensor contraction is performed on blocked matrices. In Line 15 we set the current $C_{in,im}$ block to zero if this is the first time we encounter it. Finally, lines 16-17 execute the actual tensor contraction via the BRGEMM TPP.

We highlight the separation of concerns in the heart of PARLOOPER's design: The logical loops are declared at a high-level in as little as 3 lines of code (lines 6-8 in Listing~\ref{lst:gemm} and Figure~\ref{fig:workflow}-Box A1). The main computational core is orthogonal to the loop declaration and is expressed via TPPs in 5 lines of code by making use of the \emph{logical} loop indices (Figure~\ref{fig:workflow}-Box A2). Both the loop-nest instantiation and the TPP code generation are abstracted from the user and are determined at run-time considering runtime parameters (\emph{loop\_spec\_string} for the loop generation and the target platform/problem shapes for the TPP code generation -- Figure~\ref{fig:workflow}-Box C). E.g., on a CPU platform with 3 levels of cache one may block each of the logical loops up to 3 times to maximize the data re-use out of each level of cache~\cite{goto2008anatomy}. Moreover, skewed tensor sizes may favor loop orders where the ``smaller" tensor is accessed repeatedly in the innermost loop. The parallelization of the relevant $M$ and $N$ loops also affects the data locality and the degree of data-sharing among cores. All these loop-instantiation decisions are completely determined via the runtime knob \emph{loop\_spec\_string}. Similarly, the JITed code for the BRGEMM TPP (incl. loop unrolling, vectorization, register blocking, instruction selection) depends on the target platform and the TPP shapes~\cite{georganas2021tensor}. For example, on Intel Sapphire Rapids the BF16 TPP backend can emit Advanced Matrix eXtension (AMX) instructions to accelerate the tensor contractions, whereas on an AMD Zen4 platform the TPP backend can emit AVX512 BF16 FMA instructions. All these low-level details are completely abstracted from the user-level code Figure~\ref{fig:workflow}-Box A.

Also, the GEMM in Listing~\ref{lst:gemm} is \emph{data-type agnostic}. All tensors have templated datatype, and the TPPs are \emph{precision-aware} per design: depending on how they are setup, they operate with any supported datatype internally. Therefore, the same code works for all precisions without any change.

As mentioned earlier, the legality/validity check of the \emph{loop\_spec\_string} is responsibility of
the user entity and depends on the computation at hand. An advanced framework, where the legality check is not user's responsibility, requires two steps/components: i) extracting the semantics of the loop body (written via TPPs) in a way that the data dependencies of the core computation are exposed, and ii) perform a traditional loop-dependence analysis on the loop permutation at hand to check its legality. These steps essentially would bring the current PARLOOPER/TPP framework closer to a \emph{``Tensor compiler stack with TPP abstractions"}. In the current form of the PARLOOPER/TPP framework, the correctness burden from the user’s point of view is equivalent to writing parallel code in an imperative language (e.g.\ C with OpenMP): for any nested loop with OpenMP parallelization directives, it is the user’s responsibility to dictate which loops have to be parallelized without introducing race conditions, otherwise the program is incorrect (i.e.\ the compiler does not check legality of the parallelization).

\subsection{Auto-tuning the nested loops}
\label{subsec:autotune}
A proper value of \emph{loop\_spec\_string} is essential in order to obtain the best performance on a given combination of problem size and CPU architecture. One way to select \emph{loop\_spec\_string} is to hand-craft specific loop strings that are expected to be performant given custom performance modeling of the given platform/problem (Boxes B1 and B3 in Figure~\ref{fig:workflow}). Alternatively, one may rely on off-line \emph{auto-tuning} (Box B2 in Figure~\ref{fig:workflow}). Given a logical loop declaration, along with its specifications, the tunable parameters are: (i) How many times to block each logical loops, (ii) What are the blocking sizes, (iii) Which loops to parallelize, and (iv) In which order to arrange the loops. A key observation is that all these decisions can be mapped in 1-on-1 fashion to a specific \emph{loop\_spec\_string} along with a list of block sizes.

We created an infrastructure to auto-generate an exhaustive list of strings that observe a set of constraints (see Figure~\ref{fig:workflow}-Box B2). Considering the GEMM in Listing~\ref{lst:gemm} and the decisions (i)-(iv) to be made, one can specify the following set of constraints:
\begin{enumerate}[labelindent=0.1em,labelsep=0.1cm,leftmargin=*,noitemsep,topsep=0pt]
\item Block loop \emph{a} up to 2 times, and loops \emph{b} and \emph{c} up to 3 times. This captures multi-level caches on modern CPUs.
\item For each \emph{logical} loop $i$, consider its logical trip count $T_i$ and find the prime factorization of $Ti=p_0 \cdot ... \cdot p_n$. Then pick as block factors the prefix products of the prime factors with the loop's steps. E.g., for the $M$ loop we can select as  $l0\_m\_step = m\_step \cdot p_0$ and $l1\_m\_step = m\_step \cdot p_0 \cdot p_1$. This is one example of programmatically picking the blocking factors.
\item We may decide to parallelize the $M$ (\emph{b}) and the $N$ (\emph{c}) logical loops as those define independent tasks in GEMM. We can parallelize \emph{any} of the blocked occurrence of the  $M$/$N$ loops since they also constitute independent iterations.
\item Consider all permutations of a string subject to rules 1-3.
\end{enumerate}
With such a set of constraints, we generate a set of \emph{loop\_spec\_strings} configurations to be benchmarked off-line. We implemented this infrastructure using bash scripts, but one could use other tools like Open-tuner~\cite{ansel2014opentuner}. This auto-tuning process involves 0 lines of code change in the GEMM user-code which remains the same as the one in Listing~\ref{lst:gemm}/Figure~\ref{fig:workflow}-Box A; all different loop variants dictated by the values of \emph{loop\_spec\_strings} are auto-generated and JITed by the PARLOOPER back-end.

\subsection{Performance modeling of generated loops}
\label{subsec:perf_model}
To assist the selection of the loop-instantiation \emph{loop\_spec\_string} knob, instead of relying exclusively on manual settings/exhaustive auto-tuning, we designed a lightweight, high-level performance modeling tool (Figure~\ref{fig:workflow}-Box B3). This tool takes as inputs the logical loop order along with a candidate \emph{loop\_spec\_string}, the computational core expressed via tensor-contraction BRGEMM TPPs, and few parameters modeling the target CPU, and it simulates/predicts the performance of such a setup. With this tool, one can do off-line, cross-architecture loop-tunings, and obtain a set of good loop-instantiation knobs.

Currently the POC implementation of this tool works only for PARLOOPER-generated loops with BRGEMM TPPs within \emph{body\_func} but may be extended to cover the rest TPPs. The idea is rather simple: If one were to focus on the execution of a specific thread in a specific loop-instantiation with BRGEMM TPP as main computation, all the data that are accessed are logical Tensor slices identified by the tensor-block indices. In Listing~\ref{lst:gemm}, the corresponding slices of $A_{im,ik}$, $B_{in,ik}$ and $C_{in,im}$ are characterized by the relevant indices. Consequently, each thread can create a \emph{trace} of its $A$, $B$ and $C$ accesses that arise in chronological order as the thread proceeds with the execution of the specific loop-istantiation. These traces are compact since they register accesses of full tensor slices instead of individual cache-lines~\cite{tavarageri2021polydl}. Therefore, running a data reuse algorithm that simulates multilevel caches on a per-thread basis is a low-overhead operation.

We simulate up to 3 levels of caches, each having its own size and bandwidth. The replacement policy for each cache is Least Recently Used (LRU). When processing a thread's trace access $t_i$ in chronological order, we query if the corresponding tensor slice is in any level of its cache or if it resides in memory. Then, given a BRGEMM execution at iteration $it$, and knowing the memory/cache-locations of the current slices $A_{it}$, $B_{it}$ and $C_{it}$ we can predict the execution cycles taken by BRGEMM by accounting for the relative cache bandwidths and the compute-peak of the platform. Each level of cache is represented as set and is updated based on the LRU policy as the execution progresses. For simplicity we ignore data-sharing, but the traces could be processed in lock-step fashion to account for common sub-tensors in shared levels of cache. Despite these simplifications, our performance modelling is able to single out loop instantiations with poor temporal/spatial locality, and identifies parallel schedules with poor concurrency. Such inefficient \emph{loop\_spec\_strings} are assigned a low  score and can be considered sub-optimal. %In Section~\ref{subsec:standalone_perf} we illustrate a couple of representative examples with the performance modeling tool.

\section{Kernels Developed via PARLOOPER/TPP}
We developed the following kernels via PARLOOPER/TPP that comprise the computational heart of the DL workloads presented in the next section~\ref{sec:parlooper_workloads}: (i) GEMM and Multi-Layer Perceptrons covering various tensor contractions, (ii) Convolution kernels used in computer-vision applications and (iii) Block-SpMM kernels used for efficient Transformer inference.
\label{sec:parlooper_kernels}
\subsection{GEMM and Multi-Layer Perceptron (MLP)}
\label{subsec:mlp}
\subsubsection{Extending GEMM with Activation Functions to MLP}
We extend Listing~\ref{lst:gemm} to include \emph{activation functions}, expressed via relevant TPPs. E.g., once a $C_{in,im}$ tensor block has been computed, we call a TPP on the computed 2D sub-block. In terms of the core body function, this merely translates to (for Rectified Linear Unit (RELU) activation function):
\begin{lstlisting}[style=CStyle]
if (ik == Kb-k_step) relu_tpp( &C[in][im][0][0] );
\end{lstlisting}
The GEMM kernel is used to implement a Fully-Connected Layer: an input activation tensor $I$ is multiplied with a weight tensor $W$ to yield the output tensor $O = W\times I$. A Multi-Layer Perceptrons (MLP) consists of (at least three) \emph{fully connected} layers of neurons. Each neuron in the topology may use a non-linear activation function (e.g.\ a RELU after the GEMM). MLP and Fully-Connected Layers are ubiquitous in DL, including modern Large Language Models~\cite{brown2020language} and Recommendation Systems~\cite{naumov2019deep}. An MLP within the PARLOOPER framework is just another loop around the GEMM primitive to capture the cascading GEMMs. The tensor $W_l$ of each layer $l$ corresponds to the $A$ tensor of the pertinent GEMM. Due to the cascading nature of MLP, the output matrix $O_l$ of a layer $l$ (tensor $C$ in GEMM) is subsequently the input matrix $I_{l+1}$ of the next layer $l+1$ (tensor $B$ in the next GEMM).

\subsubsection{Extending the TPP backend with AArch64 SVE support}
\label{subsubsec:mmla}
To increase the coverage of the platforms with our framework, we extend the open-source implementation of TPP in LIBXSMM~\cite{libxsmm} with support for Arm Scalable Vector Extension (SVE) ISA on AArch64 architectures~\cite{stephens2017arm}. 
Since SVE introduces predicate registers and masked operations, the SVE-TPP backend code generation follows the same strategies as the AVX256/AVX512 for x86 code generation in LIBXSMM. 

We highlight here our implementation of the (BR)GEMM TPP with SVE Matrix-Multiply Accumulate (MMLA) instructions. These instructions interprete the vector registers as packed $x\times y$ matrices. E.g., considering a 16-bit datatype (BF16) and a vector length of 256-bits (SVE256) each vector register holds 16 elements. These 16 elements are interpreted as 2 concatenated $2\times 4$ matrices. The BF16-MMLA instruction multiples the $2\times 4$ matrix of BF16 values held in each 128-bit segment of the first source vector with the $4\times 2$ BF16 matrix in the corresponding segment of the second source vector. The resulting $2\times 2$ single-precision (FP32) matrix product occupies the corresponding 128-bit segment of the output vector. The backend implementation for the (BR)GEMM TPP with MMLA instructions can follow a traditional 2D register blocking strategy~\cite{georganas2020harnessing} as long as the input tensors $A$ and $B$ are reformatted to pack $2\times 4$ and $4\times 2$ sub-tensors. The TPP collection provides the corresponding reformatting primitives. We also implement BRGEMM on SVE with MMLA support where the $B$ matrix is on-line packed to support the case where tensor $B$ is in flat format.

\subsection{Convolution Kernels}
\label{subsec:conv}
\begin{lstfloat}[t!]
\begin{lstlisting}[style=CStyle]
DType I[N][Cb][H][W][bc]; //Block C dim with bc
DType W[Kb][KCb][R][S][bc][bk]; //Block K with bk and C with bc
DType O[N][Kb][P][Q][bk]; //Block K dim with bk

auto conv_loop = ThreadedLoop<7>({
      LoopSpecs{0, N,  n_step},  // loop "a"
      LoopSpecs{0, Cb, c_step},  // loop "b"
      LoopSpecs{0, Kb, k_step},  // loop "c"
      LoopSpecs{0, P,  h_step},  // loop "d"
      LoopSpecs{0, Q,  w_step},  // loop "e"
      LoopSpecs{0, R,  r_step},  // loop "f"
      LoopSpecs{0, S,  s_step}}, // loop "g"
      loop_specs_str);

conv_loop(
  [&](int* ind) {
  int in = ind[0], ic = ind[1], ik = ind[2], ih = ind[3];
  int iw = ind[4], ir = ind[5], is = ind[6];
  unsigned long long brcount = c_step * r_step * s_step;
  if (ic == 0 && ir == 0 && is == 0) {
    zero_tpp(&O[in][ik][ih][iw][0]);
  }
  brgemm_tpp(&W[ik][ic][ir][is][0][0], 
             &I[in][ic][ih*str_h+ir][iw*str_w+is][0],
             &O[in][ik][ih][iw][0], &brcount);
});
\end{lstlisting}
%\vspace{-1.5em}
\caption{Forward convolution with PARLOOPER and TPPs}
\label{lst:conv}
\end{lstfloat}
%Here we describe the forward convolution kernel developed with PARLOOPER and TPPs.
 We employ the direct convolution method that uses the BRGEMM TPP for contractions~\cite{sc18,georganas2020harnessing}. The input tensor $I$ is convoluted with the weight tensor $W$ to yield the output $O$. The activation tensor conceptually consist of 4 dimensions: the minibatch $N$, the number of feature maps $C$ and the spatial dims $H$ and $W$. We denote the dimensions of $I$ with $N$, $C$, $H$ and $W$ while the corresponding dimensions of $O$ are $N$, $K$, $P$ and $Q$. The weight tensor is logically consisting of 4 dimensions: the feature map dimensions $C$, $K$ and the spatial dimensions $R$ and $S$. We block the dimension $C$ by $bc$ and the dimension $K$ by $bk$, and we get the blocked tensors in lines 1-3 of Listing~\ref{lst:conv}. We declare with PARLOOPER the 7 logical loops that traverse the iteration space in lines 5-13 of Listing~\ref{lst:conv}. Finally, lines 16-25 express the convolution compute kernel in a lambda function by means of the 7 logical indices and the \emph{brgemm\_tpp}. For convolutions with $R=S=1$ we can setup a stride-based BRGEMM whereas for the other cases we can setup offset-based BRGEMM to further increase the kernel's performance (i.e. the $R$ and $S$ loops would be folded in the BRGEMM by using offsets arrays~\cite{georganas2021tensor}).

\subsection{Block-Sparse $\times$ Dense Matrix Multiply (Block-SpMM)}
\label{subsec:spgemm}
\begin{lstfloat}[t!]
\begin{lstlisting}[style=CStyle]
DType A_vals[NNZ]; // non-zero values of A in BCSC
int Acolptr[M/bm], Arowidx[NNZ/(bm*bk)]//Aux structs for A in BCSC
DType B[Nb][K/v][bn][v]; // B in vnni-packed format 
DType C[Nb][M/v][bn][v]; // C in vnni-packed format

auto bcsc_spmm_loop = ThreadedLoop<3>({
  LoopSpecs{0, K,  k_step, {l1_k_step, l0_k_step}}, // K-loop "a"
  LoopSpecs{0, Mb, m_step, {l1_m_step, l0_m_step}}, // M-loop "b"
  LoopSpecs{0, Nb, n_step, {l1_n_step, l0_n_step}}},// N-loop "c"
  loop_spec_string);
  
bcsc_spmm_loop(
  [&](int* ind) {
  int ik = ind[0], im = ind[1], in = ind[2];
  if (ik == 0) zero_tpp( &C[in][(im*bm)/v][0][(im*bm)%v] );
  bcsc_spmm_tpp( A_vals, &Acolptr[im], Arowidx,
                   &B[in][ik/v][0][ik%v],
                   &C[in][(im*bm)/v][0][(im*bm)%v] );
});
\end{lstlisting}
%\vspace{-1.5em}
\caption{Block-SpMM written with PARLOOPER and TPPs}
\label{lst:spgemm}
\end{lstfloat}

The Sparse $\times$ Dense Matrix Multiplication (SpMM) is a kernel widely used in HPC, including linear solvers and graph analytics~\cite{patwary2015parallel}. The SpMM $C=A\times B$ involves one sparse matrix $A$ and two dense matrices $B$ and $C$. By exploiting the sparsity of $A$ one can reduce both the bandwidth and the compute requirements of the computation. In addition, by enforcing block sparsity on matrix $A$ one can increase the performance of SpMM by having better data locality in the accesses of the sparse $A$ and by leveraging low-precision/hardware acceleration available on modern CPU and GPU platforms. With the recent explosion in model sizes, and the intrinsic redundancy/sparsity of large models, a promising research vector is exploring block sparsity in DL~\cite{hoefler2021sparsity, lagunas2021block}.

We extend the GEMM TPP to support $A$ being block-sparse, while $B$ and $C$ matrices remain dense. We implement block sparse $\times$ dense kernel with BF16 datatype to leverage Fused Multiple-ADD (FMA) hardware-acceleration on x86 and AArch64 CPU platforms. For our implementation, we extend the TPP implementation in LIBXSMM. We enable support for $A$ in Block Compressed Sparse Columns (BCSC) format where the block-size $bm\times bk$ is parameterized. The code-generation of the microkernel works as follows: We iterate over a block row of $A$ and for each non-empty block $bm\times bk$, we multiply it with the corresponding dense block $bk\times bn$ of $B$. The dense multiplication of $bm\times bk$ with $bk\times bn$ sub-blocks of $A$ and $B$ is using 2D register blocking~\cite{georganas2020harnessing} whenever possible (i.e.\ large $bn$ and $bm$) to hide FMA latency and maximize data-reuse from registers. Aiming to deploy low-precision instructions, we opt to use a \emph{packed} format for the dense $B$ which is pre-formatted in \emph{VNNI} layout (see lines 3-4 in Listing~\ref{lst:spgemm} where $v$ is the \emph{vnni} blocking-factor). We implement this Block-SpMM TPP for x86 with both avx512-bf16 and AMX-bf16 acceleration, and for AArch64 with both BF16 dot-product and BF16-MMLA instructions.

The PARLOOPER kernel in Listing~\ref{lst:spgemm} for block-SpMM is similar to the regular GEMM. Lines 7-9 declare the loops of the sparse GEMM, while the main compute kernel is the \emph{bcsc\_spmm\_tpp} sketched in the previous paragraph. This TPP gets as input arguments the sparse matrix A in BCSC format (line 16) and matrices $B$ and $C$ are regular dense tensors. 

\section{DL Workloads via PARLOOPER/TPP}
We implemented end-to-end contemporary DL workloads via PARLOOPER/TPP within PyTorch C++ extensions in order to show the viability of our approach in real-world scenarios. We opted for the following workloads: (i) Transformer models (BERT~\cite{devlin2018bert}) to capture encoder architectures with attention mechanisms, (ii) Large Language Models (LLM GPT-J~\cite{gptj} and Llama2~\cite{llama2}) to capture decoder-only transformer architectures used in auto regressive language modeling, (iii) Sparse transformer models for efficient inference and (iv) Convolution Neural Networks (CNNs) that are prevalent in computer vision applications.

\label{sec:parlooper_workloads}

\subsection{Large Language Model Training and Inference}
\label{subsec:bert}
\begin{lstfloat}[t!]
\begin{lstlisting}[style=CStyle]
bert_output_loops(
  [&](int* ind) {
  int nc = ind[0], s1 = ind[1], nk = ind[2];
  if (nc == 0) {
    ~copy_bias_tpp~(&bias[nk], &dout[s1][nk][0]);
  }
  ~brgemm_tpp~(&in[s1][nc][0][0],
             &wt_V[nk][nc][0][0],
             &dout[s1][nk][0][0], Ncb);
  if ((nc + Ncb) < Nc) continue;
  if (p > 0) {
    ~dropout_tpp~(&dout[s1][nk][0][0], (void*)get_rng_state(),
                &dout[s1][nk][0][0], &dp_mask[s1][nk][0][0]);
  }
  ~add_tpp~(&dout[s1][nk][0][0],&in2[s1][nk][0][0],
          &dout[s1][nk][0][0]);
  if (!parallelized_on_nk && nk == (Nk - 1)) {
    ~layernorm_tpp_eqn~(&dout[s1][0][0][0],&gamma[0][0],&beta[0][0],
                      &mean[s1], &var[s1], &out[s1][0][0][0]);
  }
});
\end{lstlisting}
%\vspace{-1.5em}
\caption{Bert-Output module with PARLOOPER and TPPs}
\label{lst:bert_output}
\end{lstfloat}
 The BERT model is a transformer pre-trained via a mixture of masked language modeling objective, and next-sentence prediction~\cite{devlin2018bert}. We implemented the end-to-end workload using the architecture from Hugging Face~\cite{wolf-etal-2020-transformers}. We implemented four fused layers as PyTorch extensions in C++ using PARLOOPER and TPP: \emph{Bert-Embeddings}, \emph{Bert-Self-Attention}, \emph{Bert-Output}/\emph{Bert-SelfOutput} and \emph{Bert-Intermediate} layers. In Listing~\ref{lst:bert_output} we show an implementation of the \emph{Bert-Output}/\emph{Bert-SelfOutput} module with PARLOOPER/TPPs. We invoke a BRGEMM TPP over blocked tensor layouts, and fuse bias TPP, dropout TPP, residual add TPPs and layernorm-equation TPPs~\cite{georganas2021tensor} on a small 2D-block granularity to maximize the out-of-cache-reuse of tensors among subsequent operators~\cite{banerjee2019optimizing,zhang2018deepcpu}. Unlike previous work~\cite{georganas2021tensor} where the nested loops have a fixed ordering and parallelization, we abstract the specific instantiation with PARLOOPER and auto-tune them for the problem sizes of interest. In a similar fashion, the \emph{Bert-Self-Attention} layer consists of blocked tensor contractions, fused with scale, add, dropout and softmax TPP blocks. The \emph{Bert-Embeddings} layer is comprised of embedded lookups, followed by layernorm and dropout TPPs. Finally, \emph{Bert-Intermediate} layer is composed of BRGEMM TPP, cascaded by bias add and Gaussian Error Linear Unit (GELU) TPP. All modules are implemented via the PARLOOPER/TPP paradigm and are auto-tuned for specific shapes.

Further, by composing the aforementioned Transformer building-blocks in different ways we can build inference LLM architectures/pipelines like GPT-J~\cite{gptj} and Llama2~\cite{llama2}.

\subsection{Unstructured block-sparse BERT inference}
\label{subsec:block_sparse_bert}
We extend the BERT PyTorch extension described in Section~\ref{subsec:bert} to leverage the Block-SpMM kernel for the tensor contractions instead of dense BRGEMM TPPs. To get a block-sparse BERT model, we follow the methodology of the recent work~\cite{kurtic2022optimal}. In short, we obtain an unstructured block-sparse BERT model from a densely trained checkpoint, by applying knowledge distillation and block-wise weight pruning. We fine-tuned the block-sparse model for 40 epochs, and the final sparsity target was achieved in incremental fashion~\cite{kurtic2022optimal}.

\subsection{Convolutional Neural Networks (CNN)}
\label{subsec:rn50}
For ResNet-50~\cite{he2016deep} training we integrated our PARLOOPER CNN kernels into PyTorch as C++ extensions. We followed the architecture from the original paper~\cite{he2016deep}, where the convolution layers (see Listing~\ref{lst:conv}) are followed by batch-normalization layers (see prior work~\cite{georganas2021tensor} for batchnorm with TPPs). For the Fully Connected Layer (Listing~\ref{lst:gemm}) and the Pooling layers we simply used their respective TPP implementation (also through the PyTorch C++ extensions). %In Section~\ref{subsec:end2end_perf} we show the performance of the end-to-end ResNet-50 training pipeline.

\section{Experimental Results}
\label{sec:results}
We conduct experiments spanning machines with matrix accelerators (SPR, GVT3) to CPUs that are hybrid: 

\textbf{SPR}:A 2-socket Intel Xeon 8480+ CPU and 2$\times$ 256\,GB of DDR5-4800 memory with 8 channels per socket. Each CPU has 56 Golden Cove cores, with AVX-512 and Intel AMX technology. Each CPU has a TDP of 350W.

\textbf{GVT3}:An AWS Aarch64 Graviton 3 instance with 64 Neo-verse V1 cores featuring SVE incl. MMLA for matrix operations. Each socket has 8 channels of DDR5 memory and we determined through measurements that it is DDR5-4800. The TDP per socket is not disclosed by AWS.

\textbf{Zen4}:An AMD Ryzen 9 7950X desktop processor with 2-channel DDR5-6000 memory of 64\,GB. The CPU has 16 Zen4 cores with support for Intel AVX-512, and a TDP of 205W.

\textbf{ADL}:An Intel i9-12900K desktop processor with 2-channel DDR5-5600 memory and a capacity of 64\,GB. The CPU features 8 Intel Golden Cove Performance Cores and 8 Intel Gracemont Efficiency cores and has a TDP of 241W.

For all the experiments with our PARLOOPER/TPP framework we use for the TPP JIT code generation the LIBXSMM TPP back-end which is single threaded/serial.

\subsection{Standalone Kernel Performance}
\label{subsec:standalone_perf}
\subsubsection{GEMM and MLP standalone kernels}
\label{subsec:perf_gemm_eval}
\begin{figure}[t!]
\centering
\includegraphics[width=\columnwidth]{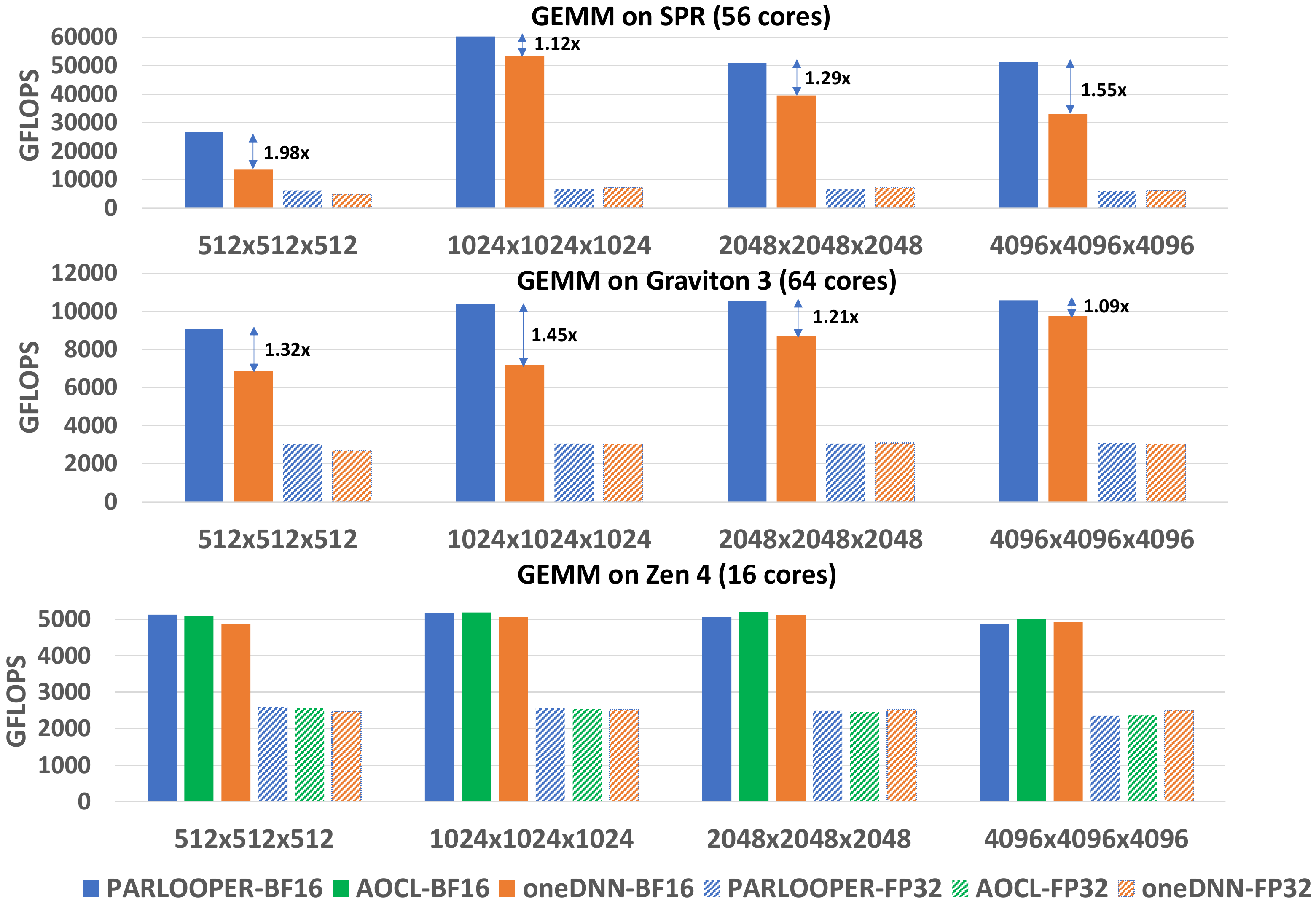}
\caption{GEMM performance of varying sizes on (Top) SPR, (Middle) Graviton 3, and (Bottom) Zen4}
\label{fig:gemm}
\end{figure}

Figure~\ref{fig:gemm} shows the performance of GEMM with varying sizes ($M\times K\times N$) on SPR (Top), GVT3 (Middle) and Zen4 (Bottom), for two precisions: FP32 (shaded bars) and BF16 (solid bars). The blue bars correspond to our implementation with PARLOOPER/TPP, while the orange bars correspond to oneDNN~\cite{onednn}. For GVT3 experiments with oneDNN, we enable the ARM Compute Library~\cite{acl} (ACL) backend which is vendor-optimized for AArch64. Our implementation with PARLOOPER and TPPs matches/exceeds the performance of the vendor-optimized libraries. While the results for FP32 are mostly on par, for BF16 we observe speedups up to 1.98$\times$ on SPR. The SPR experiments with BF16 datatype leverage the AMX instructions, offering up to 16$\times$ more peak flops than the FP32 execution with AVX512, while having access to the same memory hierarchy. Therefore, the BF16 GEMM performance is more sensitive to tensor layouts and cache blocking, since the cache/memory subsystems is stressed more due to the higher compute peak. The oneDNN implementation does not use matrix $B$ in blocked layout (see Listing~\ref{lst:gemm}) which results in extraneous cache-conflicts misses for the case with leading dimension 4096. For SPR, by leveraging BF16 and AMX we see up to 9$\times$ speedup over the FP32 execution. Similarly for GVT3, our implementation outperforms oneDNN with ACL up to 1.45$\times$ for BF16. Our newly-introduced BF16-MMLA BRGEMM kernels (Subsection~\ref{subsubsec:mmla}) offer up to 3.43$\times$ speedup over the FP32 SVE256 implementation. On Zen4 we also benchmarked the AMD-Optimized AOCL-AOCC library (version 4.1) and the results correspond to the green bars. On Zen4 all implementations perform equally well (within 4\%), and the AVX512-BF16 accelerated GEMMs observe a speedup of 2$\times$ over the FP32 versions.

\begin{figure}[t!]
\centering
\includegraphics[width=\columnwidth]{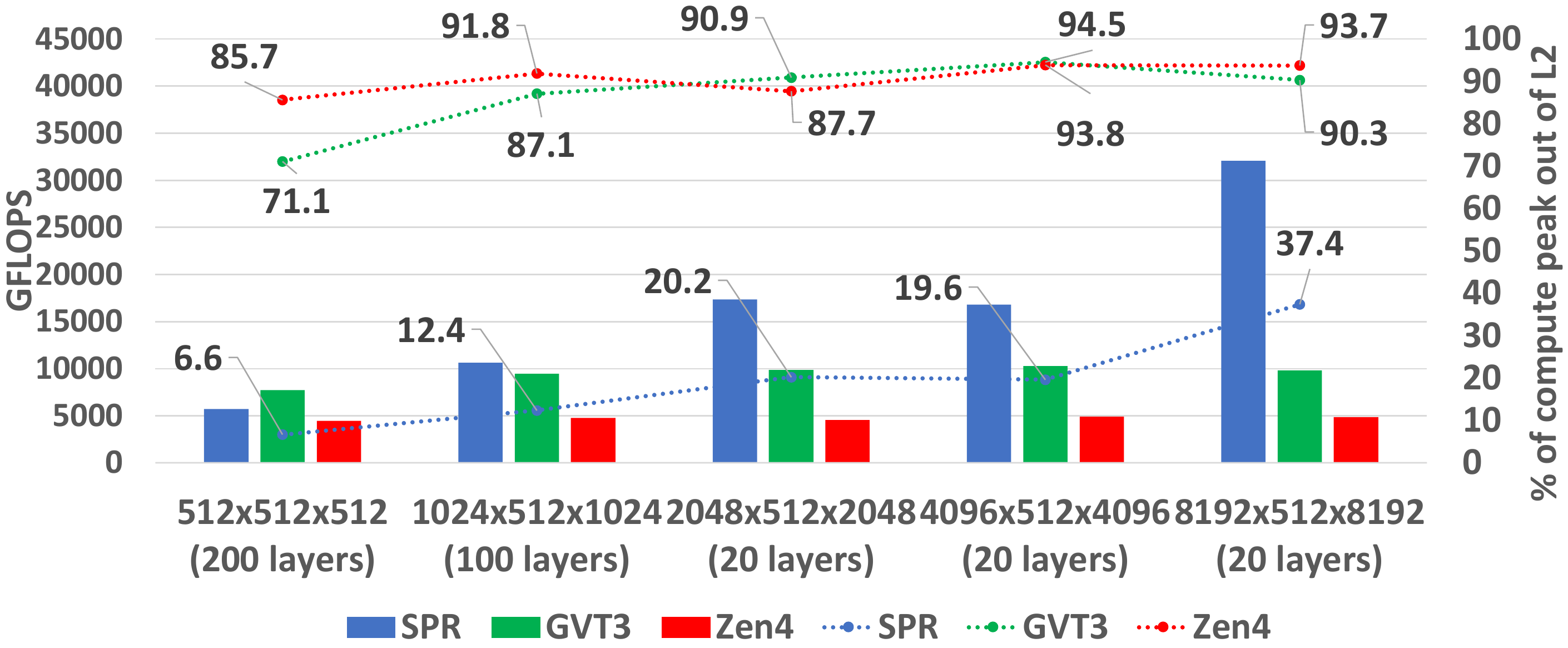}
\caption{MLP with Bias-Add and RELU activations.}
\label{fig:mlp}
\end{figure}
Figure~\ref{fig:mlp} shows the performance of the PARLOOPER BF16 MLP with Bias addition and RELU activation function. We used $N=512$ for these experiments, which corresponds to the mini-batch dimension in the DL context of MLPs, and we vary the $M$ and $K$ dimensions (i.e.\ the weight tensors of the cascading GEMMs). As we increase the weight sizes, the efficiency (dashed lines/right y-axis) keeps increasing, due to the increased $B$ tensor re-use. We observe that for SPR the efficiency maxes-out at 37.4\%. This is due to the cascading nature of activation tensors among layers, which causes core-to-core transfers as the activations flow from one layer to the next one; in such scenarios, on SPR the LLC bandwidth is the limiting factor. SPR has substantially higher compute peak than GVT3 and Zen4. Even though both the latter platforms reach more than 90\% of their compute peak, the SPR platform ends up being up to 3.3$\times$ and 6.6$\times$ faster than GVT3 and Zen4 thanks to the AMX-accelerated GEMMs.

\subsubsection{Performance comparison with TVM-Autoscheduler \& Mojo}
\begin{figure}[t!]
\centering
\includegraphics[width=1.0\columnwidth]{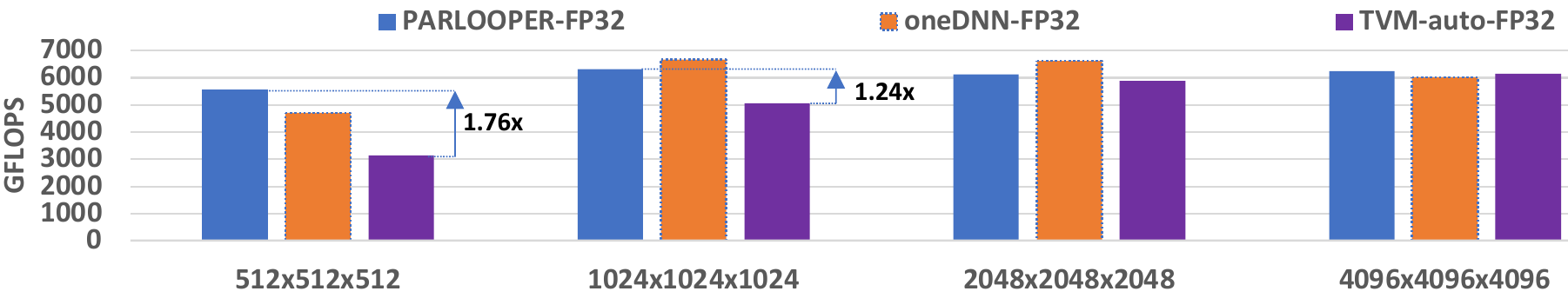}
\caption{FP32 GEMM performance on SPR with PARLOOPER, oneDNN and TVM-Autoscheduler.}
\label{fig:gemm_tvm}
\end{figure}

\begin{figure}[t!]
\centering
\includegraphics[width=1.0\columnwidth]{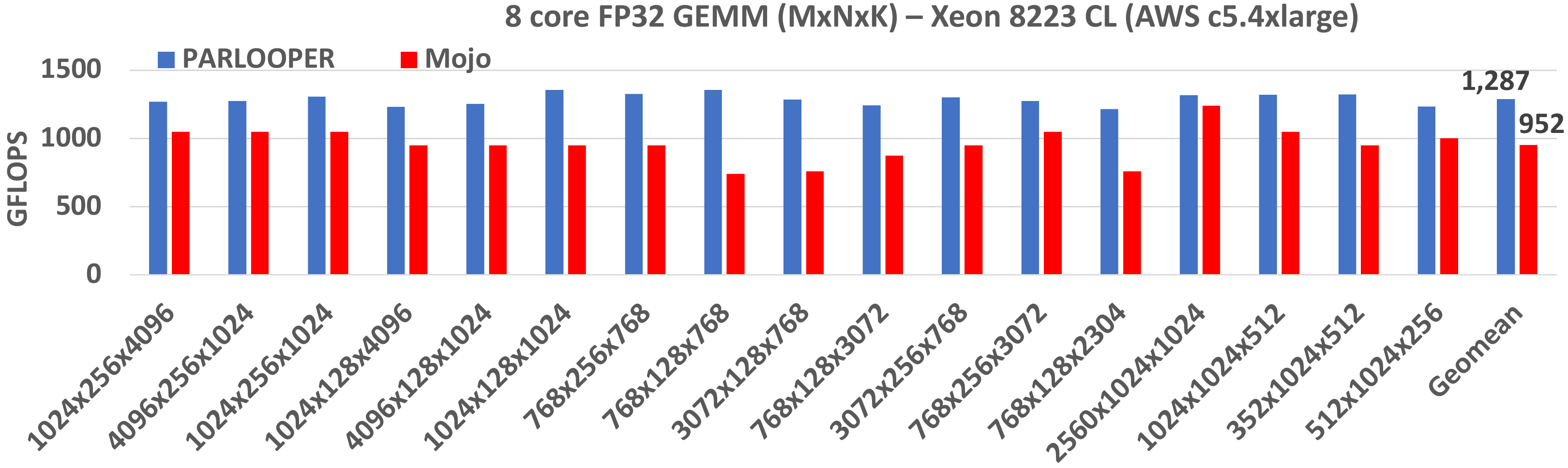}
\caption{GEMM with sizes from BERT, GPT, DLRM~\cite{modular}}
\label{fig:gemm_modular}
\end{figure}

\begin{figure}[t!]
\centering
\includegraphics[width=\columnwidth]{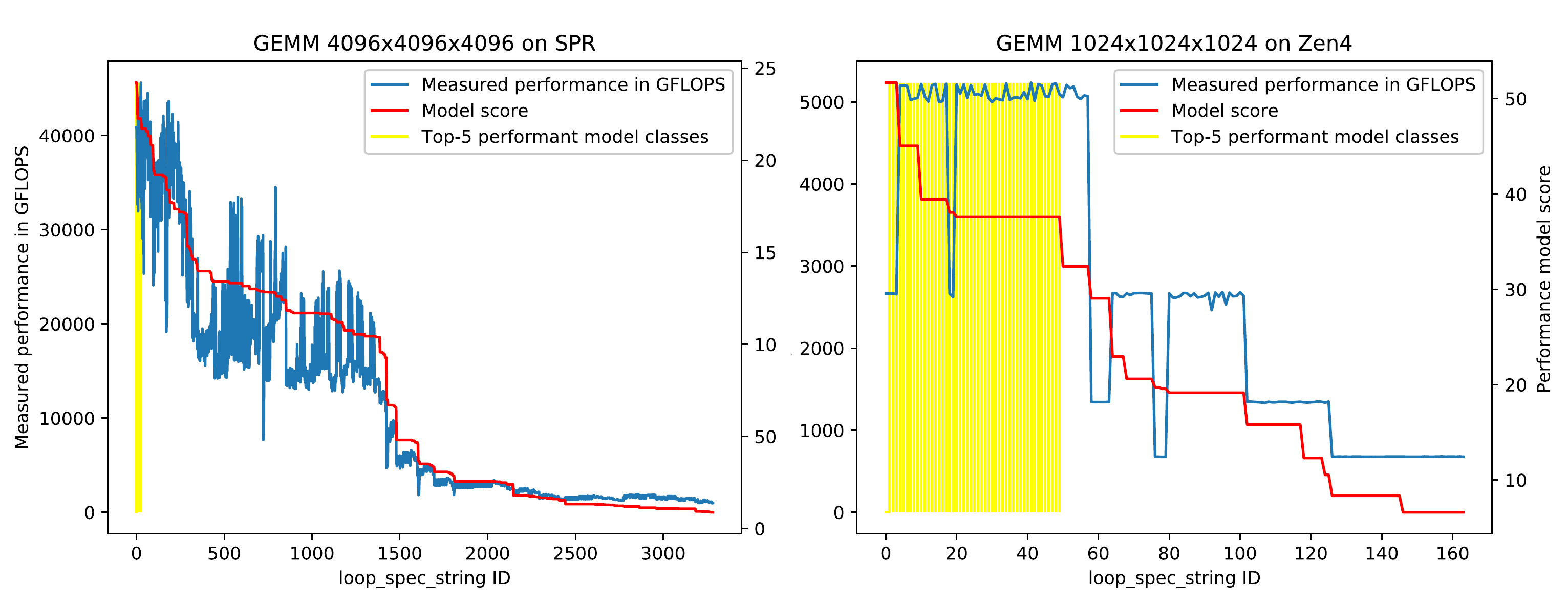}
\caption{Perf. modeling GEMMs on SPR(Left)/Zen4(Right)}
\label{fig:model}
\end{figure}

\begin{figure*}
\centering
\includegraphics[width=2.0\columnwidth]{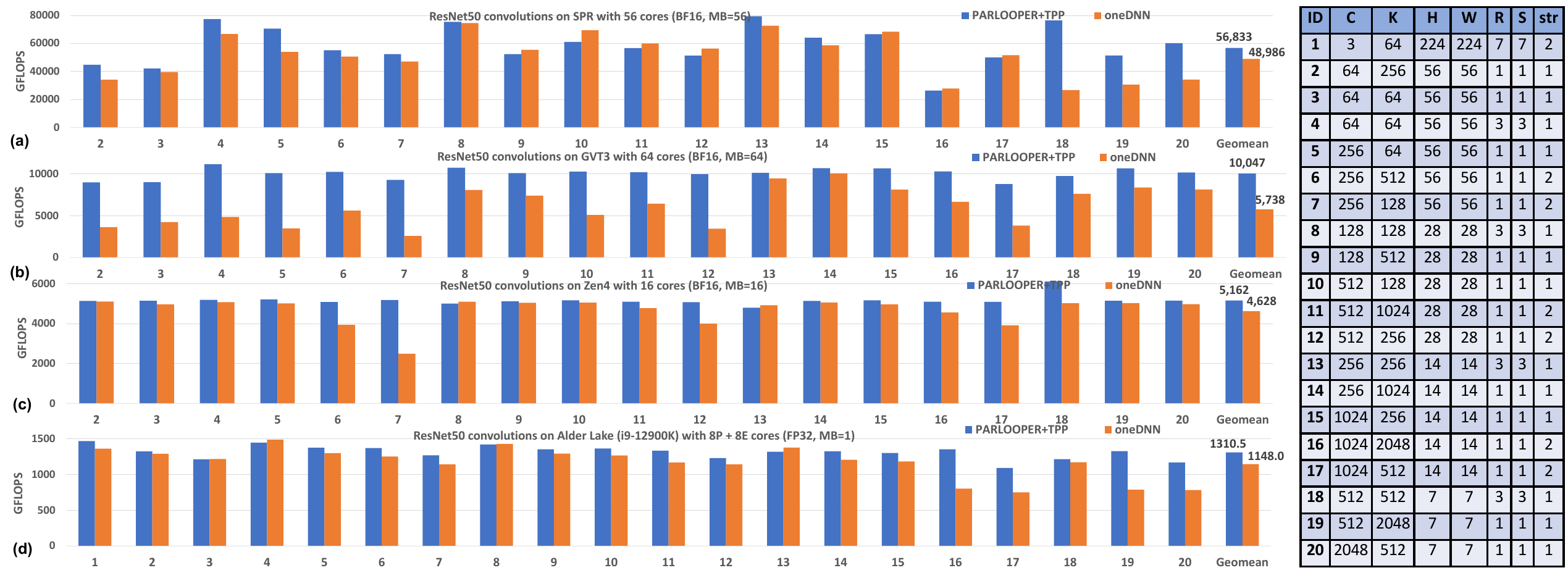}
\caption{Performance of ResNet50 convolution shapes on (a) SPR, (b) GVT3, (c) Zen4, (d) ADL. The x-axis shows the Layer ID, the y-axis shows performance in GFLOPS. Blue bars correspond to PARLOOPER/TPP, and orange bars to oneDNN.}
\label{fig:all_convolutions}
\end{figure*}

First, we compared the performance of PARLOOPER/TPP with TVM-Autoscheduler/Ansor~\cite{zheng2020ansor} which takes tensor expressions as input and generates high-performance code. We allowed tensor B to be readily reordered during the search/tuning phase (to maximize performance \& enable code generation with specialized VNNI/AMX hardware). Nevertheless, TVM-Autoscheduler was not able to generate code that leverages the hardware accelerated VNNI/AMX BF16 instructions, instead it generated slow replacement instructions sequence. This is a major limitation for practical deployment of TVM-Autoscheduler in DL workloads (also highlighted in recent work~\cite{li2023onednn}), where BF16 and low precision contractions are ubiquitous and offer speedups up to 9$\times$ over FP32 on SPR (see Figure~\ref{fig:gemm}). In Figure~\ref{fig:gemm_tvm} we compare the FP32 GEMM performance on SPR where we allowed TVM-Autoscheduler to search for 1000 schedules (recommended value from the repo). For the smaller GEMMs with limited data reuse, PARLOOPER outperforms TVM by $1.24\times$ to $1.76\times$ whereas for the larger GEMMs which are less sensitive to parallelization and tiling, TVM achieves comparable performance to PARLOOPER and oneDNN. However, the time required by TVM to obtain these schedules/kernels through search and autotuning is substantially higher than our framework. On one hand, PARLOOPER searched through $\approx$ 1000 ``outer loop" configurations for the 4 experiments in Figure~\ref{fig:gemm_tvm} in 2 seconds, 9 seconds, 2 minutes and 22 minutes respectively. On the other hand, TVM required for its autosearch/tuning 17, 18, 24 and 50 minutes respectively, being  2.3$\times$ - 500$\times$ slower than the autotuning in PARLOOPER. In the PARLOOPER/TPP framework we ``stop" the tuning/search space at the boundaries/abstraction levels of TPPs (which work at 2D-subtensor level) and as such we don’t have to search schedules/auto-tune all the way down to register-blocking/allocation/instruction selection since this level of optimizations is undertaken by the TPP backend code generation~\cite{georganas2021tensor}. On the other hand, TVM does consider the search space that pertains all the way down to vectorization, register blocking and instruction selection. Thus, the search-space in a PARLOOPER/TPP program explores solely cache blocking and parallelization options allowing it to generate up to 1.76$\times$ faster code and being up to 500$\times$ faster in the autotuning phase. These observations, in conjunction with the inability of TVM to generate performant low precision kernels make it impractical to generate fully-tuned end-to-end workloads that leverage the hardware capabilities of modern CPUs~\cite{li2023onednn,barham2019machine}.

In Figure~\ref{fig:gemm_modular} we compare the FP32 GEMM written via PARLOOPER/TPP with the GEMM written in the Mojo language~\cite{modular}. In the Mojo GEMM example, the specification of the GEMM loops is occurring at a high level along with the loop tiling, akin to PARLOOPER. However, in Mojo the user has to explicitly provide unrolling, vectorization and parallelization hints to achieve good performance. We extract the Mojo GEMM results from their blog, where the tested shapes arise from BERT, GPT, DLRM workloads, and the benchmarked CPU platform is a Xeon 8223 (an AWS c5.4xlarge instance)~\cite{modular}. We observe that the 20 LOC PARLOOPER/TPP GEMM consistently outperforms the Mojo GEMM, with a geomean speedup of 1.35$\times$.

\subsubsection{Performance modelling of GEMM}
Figure~\ref{fig:model} shows the correlation of modeling/measured results of two GEMMs on SPR (Left) and Zen4(Right) for various loops schedules/\emph{loop\_spec\_strings} (see Section~\ref{subsec:perf_model}). The blue line shows measured performance (left y-axes), and red line shows the modeled performance of the schedules (right y-axes). Our high-level performance modeling scheme is able to capture the trends pertaining to the various \emph{loop\_spec\_strings}: loops with poor locality and low-concurrency get a low-score. As a result, the top-5 modeled classes (yellow-shaded regions) always contain the most performant loop instantiation. These results indicate that our modeling tool can be used to identify performant loop-instantiation candidates. %Also, the model refinements mentioned in Section~\ref{subsec:perf_model} should be able to further improve the accuracy of the modelling. 

\subsubsection{Standalone convolution kernels benchmarking}

Figure~\ref{fig:all_convolutions} shows the PARLOOPER implementation of ResNet50~\cite{he2016deep} Convolutions. The minibatch size used on each platform equals to the number of the corresponding cores, whereas on ADL the minibatch is 1 (single-batch inference). We compare the PARLOOPER/TPP implementation (blue bars) against the oneDNN implementation (orange bars). For the first 3 platforms we benchmark BF16 convolutions (SPR, GVT3 and Zen4) whereas on ADL we benchmark FP32 since there is no BF16 hardware support on this platform. We match/exceed the oneDNN performance across platforms and convolution shapes. By considering the geometric mean on each platform (SPR/GVT3/Zen4/ADL), our implementation outperforms oneDNN by 1.16$\times$, 1.75$\times$, 1.12$\times$ and 1.14$\times$ respectively. For GVT3, the oneDNN/ACL integration is inefficient since it is using the FP32 front-end, and in the backend the input tensors are converted to BF16 on-the-fly before the actual computation with the BF16-MMLA instructions. Also, we highlight the result on ADL which used both the Performance (P) and Efficiency (E) cores. In our implementation we utilize the \emph{dynamic} OpenMP directive (see Section~\ref{subsec:loop_decl}), to account for the core heterogeneity. The convolution code with PARLOOPER/TPP across all platforms and compute-precisions is identical to the one in Listing~\ref{lst:conv}: the runtime knob \emph{loop\_spec\_string} instantiates the ``outer" loops in an optimal fashion for the problem/platform at hand, and the BRGEMM TPP backend emits optimal code for each micro-architecture. 

\subsubsection{Sparse $\times$ Dense Matrix-Multiply benchmarking}
\begin{figure}[t!]
\centering
\includegraphics[width=\columnwidth]{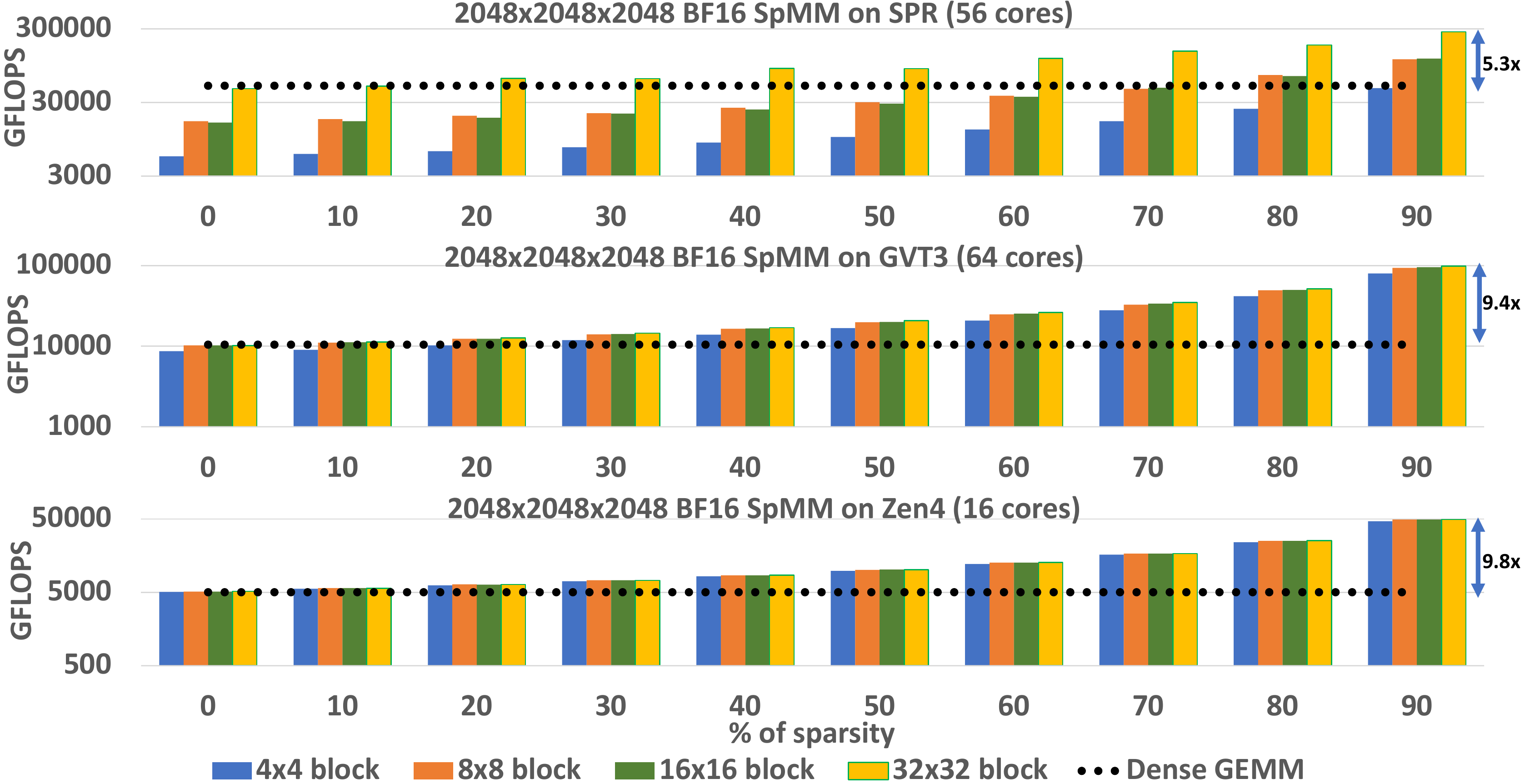}
\caption{BF16 Block-SpMM performance ($M$=$N$=$K$=2048)}
\label{fig:spmm}
\end{figure}
Figure~\ref{fig:spmm} shows the BF16 Block-SpMM performance of a 2k$\times$2k$\times$2k Matrix Multiplication on (Top) SPR, (Middle) GVT3, and (Bottom) Zen4. The x-axis shows the sparsity level, and the y-axis is in logarithmic scale dictating the \emph{effective} GFLOPS. To the best of our knowledge, there is no vendor library for low-precision block-SpMM, thus we show only PARLOOPER/TPP results. The bars correspond to various block-sizes in the block-sparsity structure. As baseline, we show with dashed line the dense GEMM performance. We see that block-sparsity together with low-precision FMA acceleration can offer speedups even for modest sparsity. On SPR and block sizes $32\times 32$ we can match the dense GEMM even without any sparsity; for 50 \% sparsity we see 1.7$\times$ speedup over the dense AMX-accelerated version, and for 90\% sparsity we see 5.3$\times$ speedup. For block sizes $16\times 16$ and $8\times 8$ we start seeing benefits for sparsity levels greater than 70\%, and for $4\times 4$ we do not see any benefits. This behavior on SPR is due to the AMX systolic array: with the larger block sizes, we are able to use efficient two-dimensional AMX tile blocking. On the contrary, for smaller block sizes, the ``dense" microkernel has small accumulation length which restricts the attainable speedup with AMX. The 4$\times$4 case is restricted to $4/32=12.5\%$ of the BF16 peak (the systolic is fully utilized with accumulation length multiples of 32). On GVT3 and Zen4 we see benefits over the dense GEMM kernels even for sparsity levels greater than 10\% for all block-sizes (the FMA-BF16 on these platforms requires accumulation chain of at least 4 and 2 respectively). For GVT3, the max attainable speedup with SpMM is 9.4$\times$ and for Zen4 it is 9.8$\times$ over their dense baselines. 

\subsection{End-to-end DL Workload Performance}
\label{subsec:end2end_perf}
\subsubsection{BERT Training Results}
\begin{figure}[t!]
\centering
\includegraphics[width=\columnwidth]{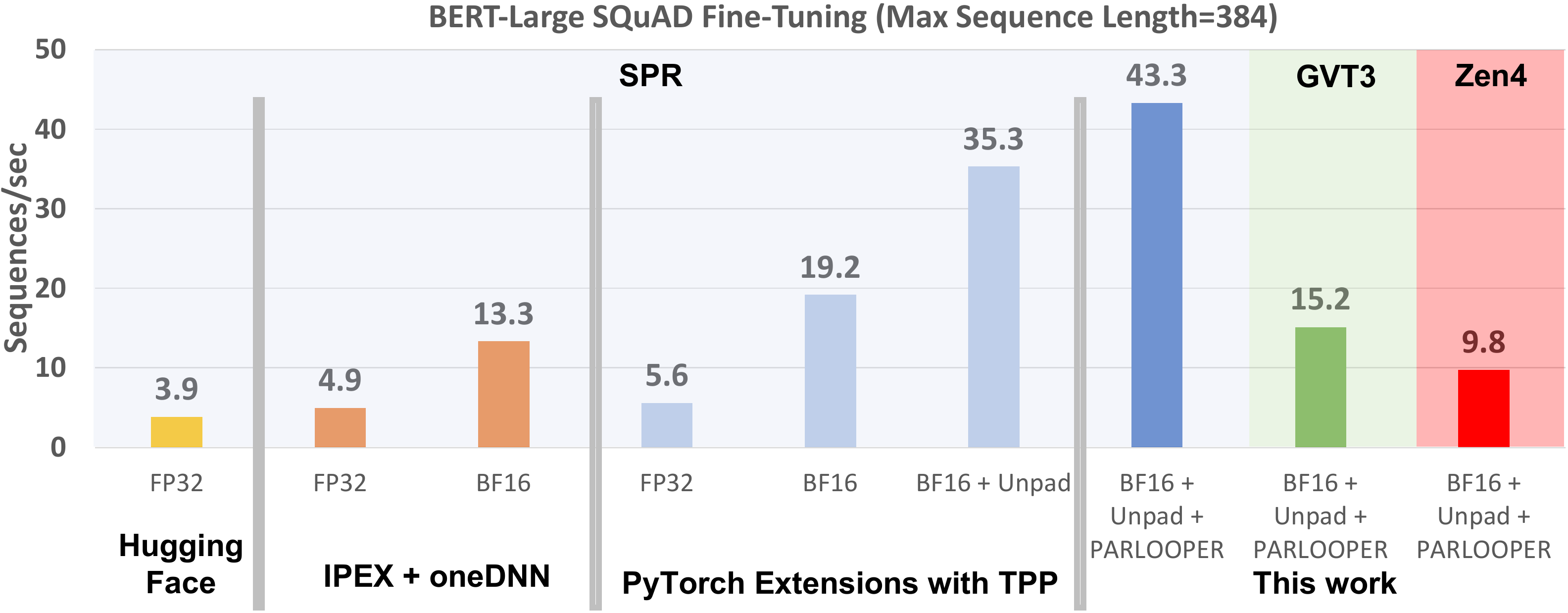}
\caption{BERT-Large SQuAD Fine-Tuning performance}
\label{fig:bert}
\end{figure}

In Figure~\ref{fig:bert} we show BERT-Large SQuAD Fine-Tuning~\cite{devlin2018bert} performance in Sequences/sec on SPR (blue-shaded area), GVT3 (green shaded area), and Zen4 (red-shaded area). On SPR we compare against alternative implementations: (i) Hugging Face (yellow)~\cite{huggingfaces}, (ii) Intel PyTorch Extensions (IPEX)~\cite{ipex} with oneDNN, and (iii) PyTorch Extensions with TPP only~\cite{georganas2021tensor}. Our implementation extends the open-source code of Georganas et al~\cite{georganas2021tensor} with PARLOOPER.

We observe that PARLOOPER/TPP speedups the SOTA implementation~\cite{georganas2021tensor} by 1.22$\times$ (43.3 vs 35.3 squences/sec). This is achieved via better loop instantiations, which we tuned specifically for the shapes of the workload and resulted in faster tensor contractions. On SPR we see tensor contraction with average performance of ~40 TFLOPS, which is in alignment with our results in Section~\ref{subsec:perf_gemm_eval} (the prior work~\cite{georganas2021tensor} merely had static loop orders in the PyTorch extensions). Our implementation, shows a speedup of 3.3$\times$ over the Intel PyTorch Extensions (IPEX)~\cite{ipex} with oneDNN, which does \emph{not} use the \emph{Unpad} Optimization that removes unnecessary computations from padded tensors~\cite{georganas2021tensor}. Our code (Listing~\ref{lst:bert_output}) is identical across all platforms, and on SPR it is 2.8$\times$ faster than GVT3 and 4.4$\times$ faster than Zen due to the higher BF16-AMX compute peak of the SPR machine. Our BERT implementation with PARLOOPER/TPP has been adopted by Intel for the recent MLPerf-v2.1~\cite{mlperf} submission (see Table~\ref{tab:mlperf_bert}) in BERT training. The SPR multi-node results (8 and 16 nodes) leverage our implementation with PARLOOPER/TPP, and is integrated with PyTorch Extensions, further highlighting the viability of our framework for real-world, distributed-memory production DL workloads. For reference, the 16-node performance is within 2.4$\times$ from the performance of 8 A100 Nvidia GPUs.

\begin{table}[t]
  \begin{center}
    \begin{tabular}{c| c}
      \textbf{System} & \textbf{Time to train (minutes)}\\\hline \hline
      8 nodes SPR (16 sockets) & 85.91\\ \hline
     16 nodes SPR (32 sockets) & 47.26\\ \hline
          DGX Box (8xA100 GPU) & 19.6\\ \hline
    \end{tabular}
  \end{center}
\caption{Time-to-train for BERT (MLPerf-v2.1-Nov.22~\cite{mlperf})}
\label{tab:mlperf_bert}
\end{table}

\subsubsection{Block-sparse BERT Inference Results}
\begin{figure}[t!]
\centering
\includegraphics[width=\columnwidth]{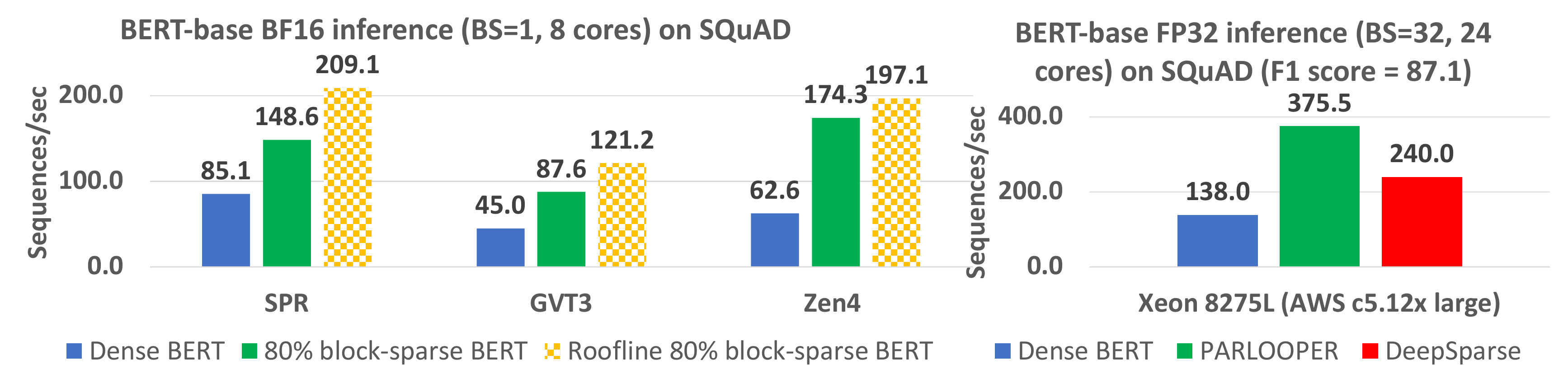}
\caption{Block-sparse BERT-Base SQuAD Inference}
\label{fig:bert_inference}
\end{figure}

In Figure~\ref{fig:bert_inference}-Left we show the inference results on the block-sparse BERT-base fine-tuned on SQuAD. We followed the methodology outlined in Section~\ref{subsec:block_sparse_bert}, and we obtained an 80\% unstructured block-sparse model with 8$\times$8 block-size. The sparse model achieves F1 score 87.1 while the dense model achieves 88.23, i.e.\ the accuracy drop is less than 1.5\%; this accuracy requirement led us to the block-size and the sparsity level of this experiment. Since we are targeting the latency-oriented BF16 inference with single Batch Size (BS=1), we use on each platform 8 cores; in practice one would execute in parallel multiple instances to fully utilize the chip. The block-sparse BERT (green bars) leveraging the block-SpMM PARLOOPER kernels achieve speedups of 1.75$\times$, 1.95$\times$, 2.79$\times$ over the dense BERT-base (blue bars) on SPR, GVT3 and Zen4 respectively. We create roofline models for each platform by assuming maximal speedup of 5$\times$ on the contractions of the workload (due to the 80\% sparsity) and the rest components do not anticipate speedup. The dotted yellow bars depict these roofline models, and our sparse BERT achieves 71\%, 72\% and 88\% of the roofline on SPR, GVT3 and Zen4 respectively.

To the best of our knowledge, there are are no vendor libraries for block-SpMM primitives with low-precision and hardware acceleration, thus we only present PARLOOPER results for BF16. In Figure~\ref{fig:bert_inference}-Right we compare our implementation with a SOTA, proprietary sparse inference runtime, namely DeepSparse~\cite{deepsparse}. We extracted the DeepSparse result from their website~\cite{deepsparse}; this experiment also corresponds to a sparse BERT-base with F1 score 87.1 (same as our 80\% block-sparse BERT-base). We used the same AWS c5.12x large instance, and the same  parameters (FP32 precision, BS=32, 24 cores) and we observe that the PARLOOPER implementation with block-SpMM is 1.56$\times$ faster than DeepSparse.

\subsubsection{Large Language Model Inference Results}
\begin{figure}[t!]
\centering
\includegraphics[width=\columnwidth]{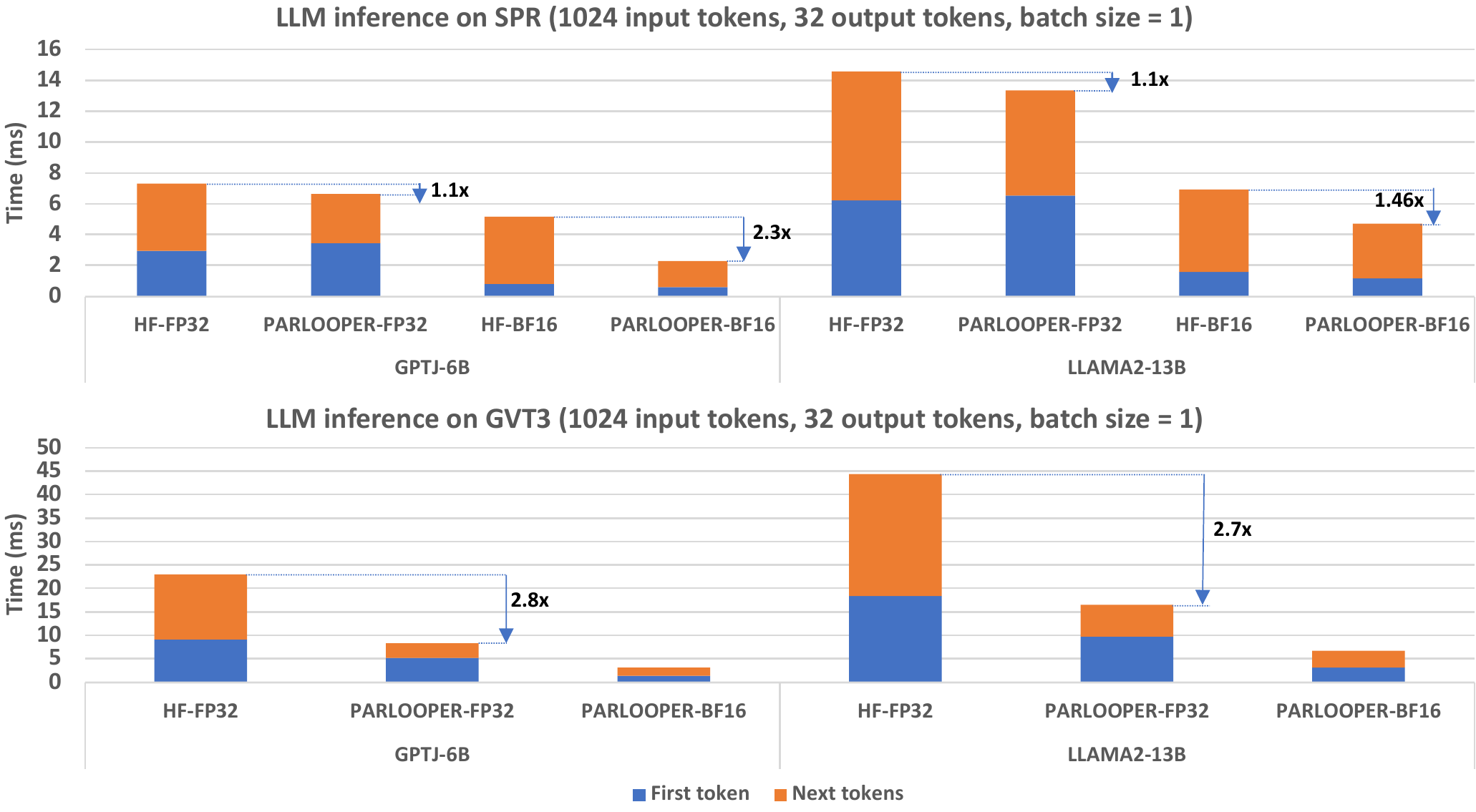}
\caption{LLM inference on SPR (Top) and GVT3 (Bottom)}
\label{fig:llm_inference}
\end{figure}
Figure~\ref{fig:llm_inference} shows the LLM inference results on SPR (Top) and GVT3 (Bottom) for GPTJ-6B~\cite{gptj} and Llama2-13B~\cite{llama2} with two software stacks: Huggingface (HF) and PyTorch extensions with PARLOOPER/TPP. All experiments use 1024 input tokens, 32 output tokens and batch size 1. On SPR, we see speedups ranging from 1.1$\times$ to 2.3$\times$ over the HF code that is using oneDNN. We also see that the BF16 datatype accelerates both the first token latency (blue portion) that is compute bound and the next tokens latency (orange portion) that is memory bandwidth bound by 5.7$\times$ and 1.9$\times$ respectively. Similarly, on GVT3 PARLOOPER is $\approx 2.8\times$ faster than the HF code. On GVT3 we note that the BF16 HF code was extremely slow and it timed out after 2 hours. We suspect this has to do with inefficient BF16 execution path of the HF code on the GVT3 platform that is using reference implementation. Nevertheless, our PARLOOPER implementation which is \emph{identical} for both platforms and all precisions was able to achieve on GVT3 $3.75\times$ and $1.84\times$ speedup by leveraging the BF16 datatype for the first and next tokens respectively.

\subsubsection{ResNet-50 training Results}
\begin{table}[t]
  \begin{center}
    \begin{tabular}{ c | c | c}
      \textbf{System} & \textbf{Implementation} & \textbf{Performance in images/sec}\\\hline \hline
      GVT3 & PARLOOPER + TPP & 145\\ \hline
      SPR & PARLOOPER + TPP & 255 \\ 
        & IPEX + oneDNN & 265\\ 
    \end{tabular}
  \end{center}
\caption{BF16 training performance of ResNet50}
\label{tab:resnet_training}
\end{table}
Table~\ref{tab:resnet_training} shows the end-to-end results for ResNet50 BF16-training on a single socket SPR and GVT3 system (Subsection~\ref{subsec:rn50}). The PARLOOPER+TPP implementation is within 4\% of the the Intel PyTorch Extensions (IPEX)~\cite{ipex} with oneDNN. The same code can be deployed on GVT3 due to the platform-agnostic nature of PARLOOPER and TPP, yielding performance within 1.76$\times$ of SPR.

\section{Related Work}
\label{sec:related}
With the rapid development of DL and HPC workloads, there is a shift to programming paradigms that break the vendor-library performance-portability ``jail". One alternative, ``library-free" approach for optimizing DL and HPC kernels is to use modern Tensor Compilers e.g.\ PlaidML~\cite{plaidml}, TVM~\cite{chen2018tvm}, Tensor Comprehensions~\cite{vasilache2018tensor}, Ansor~\cite{zheng2020ansor}, IREE~\cite{iree}, Mojo~\cite{modular}. However, current state-of-the-art tools are only capable of compiling small code blocks while large-scale operators require excessively long compilation, auto-tuning times and frequently result in code that falls short of achieving peak performance~\cite{barham2019machine}. We envision that the high-level loop and tensor abstractions we presented in this work can benefit a tensor compiler infrastructure, e.g.\ by using the TPP abstraction layer for the ISA-specific backend code generation, and by using the loop instantiation performance modeling in their internal performance models.

Domain Specific Languages (DSLs) have been traditionally used in HPC workloads, and have effectively addressed portability and performance challenges in specific scientific domains~\cite{zhang2017snowflake, ragan2013halide,portugal2016survey}. Nevertheless, these DSLs may be considered as point solutions, rather than a universal methodology across domains, and they require continuous updates to adapt to the evolving nuances of CPU architectures. To this extend, PARLOOPER could be seen as light-weight library framework for kernel development akin to DSL. However, since the loop-abstraction it provides is generic and high-level, it is not as narrow/task-specific as the majority of the DSLs.

Similar to our approach, but with applicability to GPUs is the CUTLASS~\cite{Thakkar_CUTLASS_2023} framework. It is a collection of CUDA C++ template abstractions for implementing high-performance GEMM and related computations. It decomposes the ``moving parts" of tensor contractions into reusable, modular components abstracted by C++ template classes. Primitives at different levels can be specialized and tuned via custom blocking sizes, data types, and other algorithmic strategies. All these high-level ideas are also prevalent in the design philosophy of the PARLOOPER/TPP framework. The Composable Kernel (CK) library~\cite{Liu_Composable_Kernel} aims to provide a programming model for writing performance critical kernels for machine learning workloads targeting multiple architectures (e.g.\ GPUs and CPUs). It builds on two ideas to attain performance, portability and code maintainability: (i) A tile-based programming model and (ii) Tensor Coordinate Transformation primitives. The PARLOOPER/TPP framework tackles the same problems on CPU platforms via Tensor and Loop high-level abstractions.

\section{Conclusions And Future Work}
\label{sec:conclusions}
In this work we presented a framework to develop efficient, portable DL and HPC kernels for modern CPU architectures. We demonstrate the efficacy of our approach using standalone kernels and end-to-end training and inference DL workloads that outperform state-of-the-art implementations on multiple CPU platforms. As future work, we plan to integrate the standalone kernels we developed in additional end-to-end workloads (e.g.\ DLRM~\cite{naumov2019deep}). Moreover, we plan to further apply our PARLOOPER/TPP kernel development framework on additional CPU architectures (e.g.\ with RISC-V ISA).

\pagebreak

\FloatBarrier

\bibliographystyle{unsrt}
\bibliography{references_db}

\scriptsize
\noindent
\newline Optimization Notice: Software and workloads used in
performance tests may have been optimized for performance only on
Intel microprocessors.  Performance tests, such as SYSmark and
MobileMark, are measured using specific computer systems,
components, software, operations and functions.  Any change to any
of those factors may cause the results to vary.  You should
consult other information and performance tests to assist you in
fully evaluating your contemplated purchases, including the
performance of that product when combined with other products.
For more information go to http://www.intel.com/performance.

\noindent Intel, Xeon, and Intel Xeon Phi are trademarks of Intel Corporation in the U.S. and/or other countries.

\normalsize

\end{document}